\documentclass [smallcondensed] {svjour3}
\usepackage{lettrine}
\usepackage{bm}
\usepackage{amsmath}
\usepackage{amssymb}
\usepackage{subfigure}
\usepackage{array}
\usepackage{multirow}
\usepackage[]{natbib}
\usepackage{soul}
\usepackage{booktabs}
\usepackage{placeins}
\usepackage{makecell}
\usepackage{mathtools}
\usepackage{nameref}
\usepackage{nomencl}
\usepackage[dvipsnames]{xcolor}
\usepackage{algorithm2e}
\DeclareMathOperator*{\argmin}{arg\,min}

\newcolumntype{x}[1]{>{\centering\arraybackslash}p{#1}}
\usepackage{tikz}

\makenomenclature
\begin{document}

\title{Orbit Uncertainty Propagation and Sensitivity Analysis With Separated Representations}
 
\author{Marc Balducci \and Brandon Jones \and Alireza Doostan}

\institute{Marc Balducci \at
              University of Colorado Boulder, Colorado Center for Astrodynamics Research, Boulder, CO 80309 \\
              \email{marc.balducci@colorado.edu}           
           \and
           Brandon Jones \at
              University of Colorado Boulder, Colorado Center for Astrodynamics Research, Boulder, CO 80309
              \and
              Alireza Doostan \at
              University of Colorado Boulder, Aerospace Engineering, Boulder, CO 80309
}

\date{Received: date / Accepted: date}

\maketitle

\begin{abstract}
Most approximations for stochastic differential equations with high-dimensional, non-Gaussian inputs suffer from a rapid (e.g., exponential) increase of computational cost, an issue known as the curse of dimensionality. In astrodynamics, this results in reduced accuracy when propagating an orbit-state probability density function. This paper considers the application of separated representations for orbit uncertainty propagation, where future states are expanded into a sum of products of univariate functions of initial states and other uncertain parameters. An accurate generation of separated representation requires a number of state samples that is linear in the dimension of input uncertainties. The computation cost of a separated representation scales linearly with respect to the sample count, thereby improving tractability when compared to methods that suffer from the curse of dimensionality. In addition to detailed discussions on their construction and use in sensitivity analysis, this paper presents results for three test cases of an Earth orbiting satellite.  The first two cases demonstrate that approximation via separated representations produces a tractable solution for propagating the Cartesian orbit-state uncertainty with up to 20 uncertain inputs.  The third case, which instead uses Equinoctial elements, reexamines a scenario presented in the literature and employs the proposed method for sensitivity analysis to more thoroughly characterize the relative effects of uncertain inputs on the propagated state.
\keywords{non-Gaussian \and curse of dimensionality \and sensitivity analysis \and separated representations}
\end{abstract}

\section{Introduction}

Space situational awareness (SSA) requires that an accurate estimate of a space object's state and uncertainty to be known. This estimation is a major component in identifying a possible collision and computing the probability, or reacquiring an object to facilitate a state update.  This becomes increasingly important as the ratio of objects in space to telescopes on the ground increases, which is expected with improving sensor technologies~\citep{jones}. To meet demand, SSA analyses must accurately propagate uncertainties over increasingly long time spans. Unfortunately, these involve nonlinear dynamics that cause the position and velocity multivariate probability distribution function (PDF) to become possibly highly non-Gaussian.

SSA also faces challenges in complex outer planetary and small body systems such as asteroids and comets~\citep{ryan, quadrelli}. In the case of small bodies, the gravity may be highly nonspherical and relatively unknown. Therefore if left uncontrolled for a short period, spacecraft may deviate significantly from their planned trajectories or even impact the surface~\citep{ryan}. Complex planetary systems such as the Jovian or Saturnian moon system also present SSA challenges. These environments often have many sources of perturbations. In addition to this, time between flybys of interest may be so long that the ability to utilize satellite dynamics in order to minimize fuel use is limited~\citep{quadrelli}.

{Methods of uncertainty quantification (UQ) seek to estimate the variability of a system response due to input and modeling errors.  This variability results from, for example, force model truncation and statistical uncertainties in inputs. In astrodynamics, uncertainty mapping using the state transition matrix (STM) is often used to estimate a posterior PDF~\citep{born}. As an alternative, the statistical method of MC may be used~\citep{sabol11}. Unfortunately, each method has disadvantages. The accuracy of Monte Carlo is known to be inversely proportional to the square root of the number of samples used. The result of which is that incremental improvements in the accuracy require significant increases in sample size and computation costs. However, a recent assessment of the Air Force Space Command's astrodynamic standards indicates a need for such sampling-based methods for UQ in certain scenarios~\citep{nrc_2012}. On the other hand, the STM relies on a linearization scheme, which is undesirable in the nonlinear regime of orbit propagation~\citep{junkins}. More recently, the unscented transform (UT) has been used as an efficient uncertainty propagator due to its ability to do so nonlinearly. The UT, like the STM, still assumes a Gaussian distribution and \textit{a posteriori} Gaussian assumptions have been proven to be inaccurate under certain conditions~\citep{junkins}. Cases of high variance, significant time between observations or both tend to yield non-Gaussian posterior PDFs. Reducing computation time while avoiding Gaussian assumptions in high-dimensional stochastic systems is therefore important for proposed uncertainty propagation techniques if the ratio of objects in space to sensors continues to increase.}

Methods such as Polynomial Chaos~(PC)~\citep{ghanem,xiu10a,jones,jones_mlt,jones_doostan} and Gaussian Mixture Methods~(GMM)~\citep{horwood,demars1,demars2}, which do not assume Gaussian distributions, have been proposed as an alternative to current approaches~\citep{jones}. While these new methods provide an improvement in efficiency, computation time still increases quickly (up to exponential) with respect to the number of uncertain inputs or stochastic dimensions~\citep{doostan,horwood}. This effect has been dubbed the \textit{curse of dimensionality}, the mitigation of which requires dimensional reduction or truncation, while inaction leads to increased computation time~\citep{beylkin, doostan, horwood}. State transition tensors (STT) are also being researched as a method to nonlinearly propagate state uncertainty~\citep{park, fujimoto, majji}. These STT methods, however, require the derivation of complex partial derivatives or numerical methods for their approximation.

In addition to solution statistics, it is often desired to determine the (global) \textit{sensitivity} of the solution of interest or its statistics with respect to uncertain inputs. A sensitivity analysis quantifies the relative effects that random inputs have on uncertainty in quantities of interest (QoIs)~\citep{saltelli}. The result of such an analysis can be represented via a sensitivity index based on the analysis of variance (ANOVA). These indices are helpful in a variety of circumstances, including practical (rather than mandatory) dimension truncation or prioritizing dimension determination. If the goal is to optimize reliability or minimize solution variance, among other desires, it is helpful to identify the input that is most deserving of better determination in order to reduce the output uncertainty by the largest amount possible~\citep{saltelli}. The effects of the curse of dimensionality, however, can limit the scope of sensitivity analyses by prematurely truncating stochastic dimension count for the sake of such an analysis. Although PC has been shown to efficiently compute global sensitivity indices~\citep{sudret}, the efficient generation of a PC expansion lies in the assumption of low stochastic dimension or low polynomial degree. 

As an alternative to the discussed methods, we present the approach of \textit{separated representations} (SR) for propagating uncertainties associated with the initial state of a satellite and other parameters of an orbital system to a future time. SR, also known as canonical decompositions (CANDECOMP) or parallel factor analysis (PARAFAC), provides a surrogate model to efficiently quantify the response of a system to a set of inputs. The main idea behind SR is to decompose a multivariate function of inputs into a sum of products of univariate functions of those inputs. First introduced by Hitchcock in 1927, SR has been applied to several areas including data mining, machine learning~\citep{beylkin} and chemistry~\citep{ammar,harrison}. Recent work by~\cite{beylkin05, beylkin04} shows that SR may be an effective algorithm for estimating a function of many variables, while alleviating the curse of dimensionality, i.e., SR shows promise to significantly reduce computation times for high-dimension stochastic systems~\citep{Doostan07b, Doostan09,nouy10, khoromskij10, doostan,chevreuil13,hadigol14,tamellini14,reynolds15}. SR provides a means to propagate uncertainty with nonlinear models where no assumptions of a Gaussian \textit{a posteriori} distribution are made. 

First applied to astrodynamics by~\cite{balducci}, SR has also been used to produce a direct solution of the Fokker-Planck equation for perturbed Keplerian mechanics~\citep{sun}. The expectation is that, similar to studies in the literature, computation costs of orbit UQ will decrease with respect to several other methods that fall prey to the curse of dimensionality~\citep{doostan}. {Theoretically, as demonstrated by~\cite{beylkin} and~\citep{doostan}, the number of samples needed to construct SR remains linear with respect to the number of random inputs, $d$, as far as the number of separated terms -- also known as separation rank -- is small (more precisely, independent of $d$). With this fact in mind, the available suite of random or stochastic dimensions can be expanded without significantly increasing computation time. While it is not possible to determine {\it a priori} whether or not an arbitrary QoI admits a separated representation, and in particular, with a small separation rank, it is possible to gauge the accuracy of a constructed SR from a number of additional realizations of QoI using standard statistical tools such as cross-validation \cite[Chapter 7]{friedman2001elements}. Such an {\it a posteriori} assessment of prediction accuracy has been widely used for other types of surrogate models, see, e.g., \cite[Chapter 2]{forrester2008engineering}. This, along with the linear scalability of the sample size needed for the construction of SR, makes the method competitive for UQ of high-dimensional systems.}

{
\subsection{Contributions of This Work}
This paper presents the mathematical tools required to apply SR to astrodynamics without posterior Gaussian assumptions to the PDFs. It provides convergence and regression strategies to solve for sparse solutions which admit polynomial approximations with low separation ranks. This study presents work comprised of the following contributions.

This work {includes} the first application of SR to approximating the state of a spacecraft as a function of a high-dimensional random input vector. It further illustrates the efficacy of SR in uncertainty quantification of such systems, thus encouraging future use and research in broader astrodynamics applications. Specifically, we demonstrate that SR may be used with Cartesian and equinoctial representations of the state.  By doing so, this sampling based approach for generating a surrogate model allows for use in determining the global sensitivity of the solution to its inputs. This introduces a tractable approach for future applications in mission design and optimization under uncertainty.  We also expand on methods of convergence analysis under transformations typical in astrodynamics problems, e.g., the Radial, In-Track, and Crosstrack (RIC) frame. In this later case, while the SR converges in the frame of the QoIs, additional care is required when analyzing performance under such coordinate transformations.

Leveraging the SR-based surrogate, we propose a method for performing tractable global sensitivity analyses for astrodynamic problems. The proposed method allows for the computation of global sensitivity indices in a high dimensional system, and its accuracy is compared to that of values computed with a PC methodology of known accuracy. Using this framework, we perform global sensitivity analysis of a test case presented in Horwood et al (2011), which reveals previously unknown nuances in the sensitivity of each QoIs uncertainty with respect to the random inputs.

The presentation of this work and its contributions begins with Section~\ref{sec:prbstp} introducing the setup for a general orbit problem. Section~\ref{sec:SR} then introduces the method of SR, its formulation and an implementation guideline. Following this, Section~\ref{sec:sltnstt} presents formulations for solution statistics and discusses an approach to sensitivity analysis. Section~\ref{sec:rslts} then introduces experimental tests and corresponding numeric results in order to examine the performance of SR. Of these three test cases, the first two consider the size of stochastic dimension and the resulting computation costs, while the third case explores a previously analyzed scenario and applies a sensitivity analysis. Finally, Section \ref{sec:cnclsn} concludes the paper with a summary for future work. }

\section{Problem Setup and Objective} \label{sec:prbstp}

This work considers the state of an Earth orbiting satellite. Starting with an initial condition of the QoI denoted by $\bm{q}_0 \in \mathbb{R}^M$, it is desired to know the state of the orbital system at a later time, $\bm{q}\in \mathbb{R}^M$, where $M$ denotes the dimension of the QoI. For example, when considering the Cartesian position and velocity of a satellite as QoIs, $\bm{q} = \left[x ,y ,z,\dot{x},\dot{y},\dot{z}\right]^\text{T}$ and $M = 6$. 

For the purposes of UQ, the state of the satellite is assumed to depend on $d$ random variables $\bm{\xi} \in \mathbb{R}^d$ characterizing the uncertainty in the initial state and possibly other parameters such as the coefficient of drag. The components of $\bm{\xi}$, denoted by $\xi_i$, are assumed to be independent, standard Gaussian random variables, i.e., $\bm{\xi}\sim\mathcal{N}\left(\bm 0,\bm{I}_{d\times d}\right)$, where $\bm{I}_{d\times d}$ is the $d\times d$ identity matrix. The state of the satellite at time $t$ is denoted as $\bm{q}(t, \bm{\xi})$ and satisfies a set of ODEs 
\begin{equation}\label{eq:ode}
\bm{\mathcal{F}}\left(t, \bm{\xi}; \bm{q}\right) = \bm{0}
\end{equation}
\noindent describing the temporal evolution of the satellite state. In this scenario, $t \in [t_0, t_f]$ is the temporal variable and the initial condition

\begin{equation}
\bm{q}(t_0, \bm{\xi}) = \bm{q}_0(\bm{\xi})
\end{equation}

\noindent is considered. For the interest of a cleaner presentation, the temporal dependence of $\bm{q}(\bm\xi)$ is restricted to a fixed instance of $t$ and the short notation of $\bm{q}(\bm{\xi})$ is adopted. {Our goal in this study is to approximate $\bm{q}(\bm\xi)$ as a function of $\bm\xi$, i.e., the mapping $\bm\xi\rightarrow\bm{q}(\bm\xi)$, which we will then use to estimate the statistics of $\bm{q}(\bm\xi)$, such as the mean, standard deviation (STD), and possibly marginal and joint PDFs, as well as the sensitivity of the components of $\bm{q}(\bm\xi)$ to each random input $\xi_i$.}

In the case of MC simulation, the statistics of $\bm{q}(\bm\xi)$ are estimated by calculating multiple realizations of $\bm{q}(\bm\xi)$, $\left\{\bm{q}\left(\bm{\xi}_j\right)\right\}_{j = 1}^N$, from the initial state condition $\left\{\left(\bm{\xi}_j, \bm{q}_0\left(\bm{\xi}_j\right)\right)\right\}_{j = 1}^N$, where $\bm\xi_j$ denotes an arbitrary sample of $\bm\xi$. Similarly, the $j$th sample of $\xi_i$ is denoted as $\xi_{i,j}$. In addition to moments, the joint {PDFs} of $\bm{q}(\bm\xi)$ may be approximated using a sufficiently large number of realizations of $\bm{q}(\bm{\xi})$. However, it is well-known that large values of $N$ may be needed for an accurate estimation of these statistics, thus making the method computationally intractable for certain orbit problems. Therefore, there is an interest in finding alternative methods. 

Following \cite{beylkin,doostan}, the present study examines the application of SR to approximate the mapping between $\bm\xi$ and $\bm{q}(\bm\xi)$ using random samples of $\bm\xi$, $\{\bm\xi_j\}$, and the corresponding realizations $\left\{\bm{q}\left(\bm{\xi}_j\right)\right\}$ from black box propagations. {This collection of samples and realizations is organized into the data set
\begin{equation}\label{eq:trnngDt}
 \mathcal{D} = \left\{\left(\bm{\xi}_j, \bm{q}\left(\bm{\xi}_j\right)\right)\right\}_{j = 1}^N,
\end{equation}
\noindent which is referred to as the \textit{training data}. Using a regression approach, the training data $\mathcal{D}$ is used to construct an approximation of $\bm{q}(\bm{\xi})$, where the distance between $\bm q(\bm\xi)$ and the SR approximation $\hat{\bm q}(\bm \xi)$ is minimized at the samples $\{\bm\xi_j\}$. The advantage of this approach is that the construction of $\hat{\bm{q}}(\bm{\xi})$ by regression may require far fewer realizations of $\bm{q}(\bm{\xi})$, in the form of $\mathcal{D}$, than a MC approach would need in order to accurately produce statistics or estimations of joint PDFs. By reducing the number of samples propagated in an ODE solver, the computation cost of calculating desired results is, in turn, reduced. {Although the number of samples $N$ required in $\mathcal{D}$ for an approximation is not known \textit{a priori}, the growth of $N$ with respect to stochastic dimension $d$ is linear. This relationship is discussed in further detail within Section~\ref{sec:algrthm} and can be seen in~(\ref{eq:costN}).}

The SR approximation $\hat{\bm{q}}(\bm{\xi})$ that results from this regression approach is formulated as a separable multivariate polynomial. This separated representation can then be used to analytically compute statistics of the state, or by considering the law of large numbers in a Monte Carlo fashion, it can be evaluated as a polynomial in order to derive properties such as sensitivity indices or the joint PDF. As the approximation is in the form of a multivariate function, the method of SR is able to leverage the relatively low computation cost of polynomial evaluations in order to produce results which rely on large numbers of MC realizations.}

To simplify the rest of the presentation, we define the data-dependent (semi-) inner product of two vectors $\bm q(\bm\xi),\bm r(\bm\xi)\in\mathbb{R}^{M}$ as
\begin{equation}
\label{eq:semi_inner}
\left\langle \bm q,\bm r\right\rangle_{D}=\frac{1}{N}\sum_{j=1}^{N}\left\langle \bm q(\bm\xi_j),\bm r(\bm\xi_j) \right\rangle_2,
\end{equation}
where $\left\langle \cdot,\cdot\right\rangle_{2}$ denotes the standard Euclidean inner product. The semi (semi-) inner product in (\ref{eq:semi_inner}) induces the (semi-) norm 
\begin{equation}
\label{eq:semi_norm}
\Vert\bm q\Vert_{D}=\left\langle \bm q,\bm q\right\rangle_{D}^{1/2},
\end{equation}
which is used frequently hereafter.

%
%

\section{Separated Representation} \label{sec:SR}

The separable approach is based on approximating a multivariate scalar function $q\left(\bm{\xi}\right)$ with a sum of products of univariate functions. Consider the example function in Section~\ref{sec:prbstp}, which is presented as the state of an orbiting satellite. {In the scalar setting, let the QoI $q(\bm\xi)$ be a single element of the satellite state such as the $x$-position. In the framework of SR, the separated approximation of the satellite state is then a sum of separable products
\begin{align} \label{eq:SR_scalar}
q(\bm{\xi}) \approx \hat{q}(\bm{\xi}) = \sum^r_{l = 1}{s^l} \prod^d_{i = 1} {u^l_{i}} \left(\xi_i\right),
\end{align}
where $\hat{q}(\bm{\xi})$ is the estimation  of $q(\bm\xi)$, and $\{u_i(\xi_i)\}_{i=1}^d$, are unknown univariate functions to be determined so that $\hat{q}(\bm\xi)$ is as close as possible to $q(\bm\xi)$.} Additionally, $\{u^l_{i}(\xi_i)\}_{l=1}^{r}$, $i=1,\dots,d$ are referred to as {\it factors}, and $\{s^l\}_{l=1}^{r}$ are normalization constants such that each $u^l_{i}(\xi_i)$ has a unit norm, as elaborated in Section~\ref{sec:als}. These normalization constants provide numerical stability to the formulation of an SR. {In (\ref{eq:SR_scalar}), the {\it a priori} unknown constant $r$ is referred to as the {\it separation rank} and is ideally the smallest number of separated terms -- to be determined in the construction of SR -- in order to achieve a desired accuracy in approximating $q(\bm\xi)$. The approximation $\hat{q}(\bm\xi)$ is considered to be \textit{low-rank} if $r$ remains small for the target accuracy. It is worthwhile highlighting that the separated representation (\ref{eq:SR_scalar}) is a nonlinear approximation of $q(\bm\xi)$ with small number of {\it parameters} in which the expansion basis functions $\prod^d_{i = 1} {u^l_{i}} \left(\xi_i\right)$ are not predefined; they are sought such that the approximation error is minimized, as discussed in the following. When $q(\bm\xi)$ admits a low separation rank $r$, this allows a fast decay of the error with respect to $r$. In addition, as we shall explain in Section \ref{sec:als}, the nonlinear approximation (\ref{eq:SR_scalar}) may be computed using multilinear approaches due to its separated form with respect to variables $\xi_i$. The combination of these two attributes of SR will allow its construction with a number of samples that is linear in $d$, which is smaller than that of standard approximation techniques relying on {\it a priori} fixed bases. For scenarios where the desired accuracy may not be achieved by a small $r$, e.g., when $q(\bm\xi)$ is discontinuous in $\bm\xi$ along an arbitrary hyperplane, SR may not lead to an efficient approximation. Therefore, in practice, it is crucial to assess the quality of a constructed SR -- see Remark \ref{rem:rank} and Section \ref{sec:rslts} -- prior to using it to learn an arbitrary $q(\bm\xi)$ and its statistics. Such an assessment is also a key step in the construction of other surrogate models. We refer the interested reader to the review manuscripts \cite{Kolda09b,chinesta2011short} and the references therein for examples of successful application of SR to various types of problems in engineering and sciences.} 

In details, the construction of the SR in (\ref{eq:SR_scalar}) may be posed in the form of a nonlinear optimization involving unknowns $\{u_i^l(\xi_i)\}$ in order to minimize the distance between $q(\bm\xi)$ and $\hat{q}(\bm\xi)$ at the samples $\{\bm\xi_j\}$, 
\begin{equation}
\label{eqn:min_scalar}
\min_{\{u_i^l(\xi_i)\}} \left\Vert q-\hat{q}\right\Vert_D^2,
\end{equation}
as detailed in Section~\ref{sec:als}. These unknowns are often approximated in an {\it a priori} selected basis, thus allowing for a numerical solution to the optimization problem. Here, we consider the approximation of each factor in a basis of Hermite polynomials in $\xi_i$, 
\begin{eqnarray}\label{eq:u_sclr}
{u^l_{i}} \left(\xi_i\right) &\approx& \sum_{p=1}^P{{c}^l_{i,p}}\psi_{p}(\xi_i),
\end{eqnarray}
where $\psi_{p}(\xi_i)$ is the Hermite polynomial of degree $p-1$. In general, the basis functions $\psi_{p}(\xi_i)$ are selected such they are orthogonal with respect to the PDF of $\xi_i$, as given by the Askey family of orthogonal polynomials,~\citep{askey85}. This allows for selecting a different basis for each direction $i$, depending on the distribution of $\xi_i$. Given the discretization of factors in (\ref{eq:u_sclr}), the optimization problem (\ref{eqn:min_scalar}) is now reduced to finding the unknown coefficients $\bm{c}^l_i = [c^l_{i,1}, \ldots, c^l_{i,P}]$ for each direction $i$ and rank $l$ via problem
\begin{equation}
\label{eqn:min_scalar_c}
\min_{\{\bm{c}^l_i\}} \left\Vert q-\hat{q}\right\Vert_D^2.
\end{equation}
{The number of unknowns in the optimization problem (\ref{eqn:min_scalar_c}) is $r\cdot d\cdot P$, which is linear in the dimension $d$, as far as the separation rank $r$ is independent of $d$. As discussed in \cite{beylkin}, this linear scaling of the number of unknowns in turn suggests a linear (in $d$) growth requirement on the number of samples $N$ used to solve (\ref{eqn:min_scalar_c}).} \\


{
\begin{remark}\label{rem:rank} While increasing the separation rank $r$ improves the accuracy of SR, it is not possible to determine {\it a priori} if an arbitrary $q(\bm\xi)$ lends itself to an accurate separated representation with low separation ranks. In practice, the accuracy of a constructed SR, $\hat{q}(\bm\xi)$, may be assessed empirically via, for instance, cross-validation techniques commonly used for other types of surrogate models, e.g., based on multivariate polynomial expansions (See \cite{forrester2008engineering}). For certain classes of problems, e.g., linear elliptic, \cite{Cohen10a}, semi-linear elliptic, \cite{Hansen12}, and parabolic, \cite{Hoang13}, partial differential equations with random inputs, {\it a priori} error estimates have been derived, demonstrating the sparsity of the solution when expanded in multivariate polynomial bases. That is, only a small fraction of the basis functions have non-negligible coefficients. When each SR factor $u^l_{i}(\xi_i)$ is approximated in the same (univariate) polynomial basis, as in (\ref{eq:u_sclr}), such sparse solutions are in principle guaranteed to admit SRs with low separation ranks. This is because the two approaches employ basis functions spanning the same space within which the solution exists or is well approximated.
\end{remark}
}

\subsection{SR of Vector-Valued Functions}

In many cases, it is desirable to approximate a vector-valued QoI via SR. For instance, in the current application to orbit uncertainty propagation, an estimate for the Cartesian components of position and velocity is needed. In such cases, the SR approximation of $\bm q(\bm \xi)\in\mathbb{R}^{M}$ is given by 
\begin{equation}\label{eq:SR_vec}
\bm{q}(\bm{\xi})\approx\hat{\bm{q}}(\bm{\xi}) = \sum_{l = 1}^r {s^l} \, \bm{u}_0^l \,\prod^d_{i = 1} {u^l_{i}} \left(\xi_i\right),
\end{equation}
where the definitions and approximation of $u^l_{i}(\xi_i)$ remains the same as in the scalar-valued SR. A significant difference between (\ref{eq:SR_vec}) and (\ref{eq:SR_scalar}) is the addition of the vector of deterministic factors $\bm{u}_0^l = [u^l_{0,1},\ldots,u^l_{0,M}]^{\text{T}}\in\mathbb{R}^M$, which is used to extend the approximation from the scalar-valued $q(\bm\xi)$ to the vector-valued $\bm q(\bm \xi)$ by solving the optimization problem
\begin{equation}
\label{eqn:min_vec_c}
\min_{\{\bm{c}^l_i\},\{\bm{u}_0^l\}} \left\Vert \bm{q}-\hat{\bm{q}}\right\Vert_D^2.
\end{equation}
In the case of (\ref{eq:SR_vec}), the set of scalars $s^l$ are now normalization constants such that both $u^l_{i}(\xi_i)$ and $\bm{u}_0^l$ are normalized to have unit norm. The method of determining these values is detailed in Section~\ref{sec:als}.

\subsection{Process of Constructing SR}\label{sec:apprx}

As an introduction to the implementation of SR, a high level algorithm for the vector output is provided in this section. The sampling approach to constructing SR requires the generation of training data $\mathcal{D}$, i.e., the set $\{(\bm{\xi}_j,\bm{q}(\bm{\xi}_j))\}$ of $N$ samples from Eq.~(\ref{eq:trnngDt}), that represent propagated realizations of $\bm{q}(\bm\xi)$  at (random) samples of inputs $\bm\xi$. Such algorithms are dubbed \emph{non-intrusive} as the evaluation of $\bm{q}(\bm\xi)$ does not require any alteration of the solvers for (\ref{eq:ode}). For the case of orbit state propagation, the non-intrusive SR process may be summarized by:
\begin{enumerate}
\item Generate a set of independent, random realizations $\{\bm{\xi}_j\}^N_{j=1}$, where each $\bm{\xi}_j\sim\mathcal{N}\left(\bm 0,\bm{I}_{d\times d}\right)$.
\item Using the \emph{a priori} state distribution, generate the set of samples $\{\bm{q}_0(\bm{\xi}_j)\}^N_{j=1}$ at the epoch time. 
\item Propagate each of the $N$ samples to the time of interest using a given black box ODE solver to get $\{\bm{q}(\bm{\xi}_j)\}_{j=1}^N$.
\item Use the training data $\{(\bm{\xi}_j,\bm{q}(\bm{\xi}_j))\}_{j=1}^{N}$ to generate the SR approximation.
\end{enumerate}

\noindent { Steps 1 through 3 of Section~\ref{sec:apprx} include the generation of $\mathcal{D}$. The data used in the numerical results of this paper assume an {\it a priori} Gaussian PDF, i.e., $\bm{q}_0\sim \mathcal{N}(\bar{\bm{q}}_0,\bm{\Sigma})$, with a given mean vector $\bar{\bm{q}}_0$ and covariance matrix $\bm{\Sigma}$. Each sample $\bm{\xi}_j$ is first mapped to an initial sample $\bm{q}_0(\bm{\xi}_j)$ via a Cholesky decomposition of $\bm{\Sigma}$.
The elements of $\bm{q}_0\left(\bm{\xi}_j\right)$ are then propagated using a desired integration method as a black box to produce the set $\{\bm{q}\left(\bm{\xi}_j\right)\}$ at some time of interest. }

\subsection{Solution via Alternating Least Squares}\label{sec:als}

Step 4 in Section~\ref{sec:apprx} is the generation of the SR approximation, which here is done via alternating least squares (ALS) regression using a set of $N$ training samples $\{(\bm{\xi}_j,\bm{q}(\bm{\xi}_j))\}$. This method reduces the larger nonlinear optimization process into a series of linear least squares regression problems~\citep{beylkin}. 

{\it Overall approach.} Given an initial $r$, e.g., $r=1$, pre-selected basis $\{\psi_p(\xi_i)\}$, and initial coefficient values $\{\bm{c}^l_{i}\}$ and $\{\bm{u}_0^l\}$, the ALS algorithm updates the coefficients $\{\bm{c}^l_{i}\}$ by alternating through a sequence of one-dimensional and linear optimization problems. Let $k=1,\dots,d$ denote one direction of interest for each of these problems. The coefficients $\{\bm{c}^l_{k}\}$ are updated by solving the linear least squares regression
\begin{equation}
\label{eqn:min_vec_c_k}
\{\bm{c}^l_{k}\}_{l=1}^{r} = \argmin_{\{\bm{c}^l_{k}\}_{l=1}^{r}}\ \left\Vert \bm{q}-\hat{\bm{q}}\right\Vert_D^2.
\end{equation}
while the coefficients $\{\bm{c}^l_i\}$ for other directions $i \neq k$ and $\{\bm{u}_0^l\}$ are fixed at their current values. 
%
The ALS algorithm continues with a sweep through each direction $k$ in alternation. After reaching the final direction $d$, the values of $\{\bm{u}_0^l\}$ are solved for (completing what will be referenced as a full ALS loop) using the linear least squares problem
\begin{equation}
\label{eqn:min_vec_u_k}
\{\bm{u}_0^l\}_{l=1}^{r} = \argmin_{\{\bm{u}_0^l\}_{l=1}^{r}}\ \left\Vert \bm{q}-\hat{\bm{q}}\right\Vert_D^2, 
\end{equation}
while fixing the coefficients $\{\bm{c}^l_i\}$ for all $i$ at their current values. Full loops of ALS are continued until an {\it a priori} selected convergence criteria is met. Upon this convergence of the solution with rank $r$, the SR solution is tested for solution precision when compared to the training data. If this second convergence criteria is not met, then the separation rank is increased ($r=r+1$) and the ALS procedure is repeated to generate a solution for the larger set of coefficients $\{\bm{c}^l_{i}\}$ and $\{\bm u_0^l\}$. 

The remainder of this section provides a more detailed description of the ALS method used to generate an SR approximation. \\[-.3cm]

{\it Updating stochastic coefficients $\{\bm{c}^l_k\}$.} Setting the derivative of $\left\Vert \bm{q}-\hat{\bm{q}}\right\Vert_D^2$ with respect to $\{\bm{c}^l_k\}$ to zero leads to the normal equation
\begin{equation}\label{eq:ls}
\left( \bm{A}^\text{T}\bm{A}\right)\bm{z} = \bm{A}^\text{T}\bm{h}
\end{equation}
for the solution $\{\bm{c}^l_k\}$ to problem (\ref{eqn:min_vec_c_k}), organized in 
\begin{equation}
\bm{z} = \begin{bmatrix}\left(\bm{c}^1_k\right)^\text{T} & \cdots & \left({\bm{c}^r_k}\right)^\text{T} \end{bmatrix}^\text{T}.
\end{equation}
In (\ref{eq:ls}), 
\begin{align}\label{eq:blockA}
\bm{A} = \begin{bmatrix}{\bm{A}_{11}}&\cdots&{\bm{A}_{1r}}\\ \vdots&\ddots&\vdots\\ {\bm{A}_{N1}}&\cdots&{\bm{A}_{Nr}}\end{bmatrix},
\end{align}
where the $(j,l)$ block of $\bm A$, ${\bm{A}_{jl}}\in\mathbb{R}^{M\times P}$, is given by
\begin{align}\label{eq:blocks}
{\bm{A}_{jl}} = s^l\begin{bmatrix}{\bm{u}_0^l}\,\psi_{1}(\xi_{k,j}) &\cdots&{\bm{u}_0^l}\,\psi_{P}(\xi_{k,j})\end{bmatrix}\prod_{i \neq k} {u^l_{i}} \left(\xi_{i,j}\right)
\end{align}
and the data vector 
\begin{equation}\label{eq:X_matrix}
\bm{h} = \begin{bmatrix}\bm{q}(\bm{\xi}_1)^\text{T}& \cdots & \bm{q}(\bm{\xi}_N)^\text{T} \end{bmatrix}^\text{T}\in\mathbb{R}^{MN},
\end{equation}
contains the samples of $\bm q(\bm\xi)$. Before continuing to the next direction in the alternation, the values of $s^l$, and therefore ${\bm{c}}^l$, are updated via
\begin{equation}\label{eq:s_update}
\bm{c}^l_k \leftarrow \frac{\bm{c}^l_k}{\Vert u^l_k\Vert_D}\quad\text{and}\quad {s^l} \leftarrow {s^l}\Vert u^l_k\Vert_D.
\end{equation}

{
\begin{remark} Notice that, while the total number of unknowns in (\ref{eqn:min_vec_c_k}) is $r\cdot d \cdot P$, only $r \cdot P$ of them appear in the one-dimensional optimization problems (\ref{eq:ls}), which is independent of $d$, as long as $r$ does not depend on $d$. 

\end{remark}
}

{\it Updating deterministic factors $\{\bm{u}_0^l\}$.} After cycling through all directions $k = 1, \ldots ,d$ in the ALS process to compute $\bm{c}^l_{k}$, the best estimate of $\{\bm{u}_0^l\}$ is found by setting the derivative of $\left\Vert \bm{q}-\hat{\bm{q}}\right\Vert_D^2$ with respect to $\bm{u}_0^l$ to zero. The resulting normal equation associated with problem (\ref{eqn:min_vec_u_k}) is 
\begin{equation}\label{eq:ls_det}
\left( \bm{A}^\text{T}\bm{A}\right)\bm{Z} = \bm{A}^\text{T}\bm{H},
\end{equation}
where,
\begin{align}\label{eq:blockA_tld}
\bm{A} = \begin{bmatrix}\bm{A}_1 & \cdots & \bm{A}_N \end{bmatrix}^\text{T} \in\mathbb{R}^{N\times r}
\end{align}
and each block $\bm{A}_j \in \mathbb{R}^{1 \times r}$ is of the form
\begin{equation}\label{eq:blocks_tld}
\bm{A}_j =\begin{bmatrix}{s^1} \prod\limits^d_{i = 1} {u^1_{i}} \left(\xi_{i,j}\right) && \cdots && {s^r} \prod\limits^d_{i = 1} {u^r_{i}} \left(\xi_{i,j}\right)\end{bmatrix}.
\end{equation}
Additionally, the solution and data matrices $\bm H$ and $\bm Z$ are, respectively,

\begin{align}\label{eq:h_utld}
\bm{H} = \begin{bmatrix}\bm{q}(\bm{\xi}_1)&& \cdots && \bm{q}(\bm{\xi}_N) \end{bmatrix}^\text{T}\in\mathbb{R}^{N\times M}
\end{align}
and
\begin{equation}\label{eq:z_utld}
\bm{Z} = \begin{bmatrix} \bm{u}_0^1 && \cdots && \bm{u}_0^r \end{bmatrix}^\text{T} \in\mathbb{R}^{r\times M}.
\end{equation}
Once $\bm{Z}$ is solved for using (\ref{eq:ls_det}), $\bm{u}_0^l$ is found an normalized using a method similar to (\ref{eq:s_update}),
\begin{equation}\label{eq:u_tilde_update}
\bm{u}_0^l \leftarrow \frac{\bm{u}_0^l}{\left\|\bm{u}_0^l\right\|_2}\qquad\text{and}\qquad {s^l} \leftarrow {s^l}\,\left\|\bm{u}_0^l\right\|_2.
\end{equation}
After $\{\bm{u}_0^l\}$ has been estimated and normalized, and if the solution fails the criteria for convergence, then either another rank is added or the alternation through the directions continues. As discussed in~\cite{beylkin}, the error $\left\Vert \bm{q}-\hat{\bm{q}}\right\Vert_D$ can never increase throughout these ALS updates. \\[-.3cm]


{\it Rank increase and solution convergence.} In order to develop an SR estimate that achieves desired accuracies, a set of tolerances should be defined. One must consider the minimization of error $\|\bm{q} - \hat{\bm{q}}\|_D$ between the surrogate and training data
as well as the accuracy improvement from one full ALS sweep to the next. In order to enforce a maximum $r$, the user could set the desired maximum rank or aim for a particular solution precision. The method used in this paper sets a desired maximum rank, but a precision based solution achieves convergence when the relative residual 
\begin{equation}\label{eq:conv}
\gamma = \frac{\left\|
\bm{q} - \hat{\bm{q}}
\right\|_D}{\left\|\bm{q}
\right\|_D }<\epsilon,
\end{equation}
in which $\epsilon$ is a desired relative tolerance. Theoretically, the approximation developed from SR should decrease $\gamma$ as the separation rank $r$ is increased. However, there is a limit to the precision that can be reached with a fixed rank~\citep{beylkin}. The ALS process is repeated until the difference between the surrogate and training data no longer changes significantly from one iteration to the next. If the ALS process has converged on a solution, but the minimum relative residual has yet to be reached, another rank may be added. In order to identify such a case, the current implementation uses the difference in relative residuals~(\ref{eq:conv}) as an indication of convergence. Specifically, the rank is increased when the solution convergence has not been met and the difference between the relative residual from the most recent iteration and the relative residual from two iterations previous is below some desired relative tolerance $\delta$. This process is able to determine when improvement in the precision of the surrogate stalls or becomes insignificant.
\begin{remark} We highlight that choosing a small $\epsilon$ (or an unnecessarily large $r$) may result in {\it over-fitting}; that is, the difference between $\bm q(\bm \xi)$ and its SR approximation $\hat{\bm q}(\bm \xi)$ may be large, while $\gamma$ is small. To avoid this issue and as discussed in \cite{doostan}, the least squares problems (\ref{eq:s_update}) and (\ref{eq:ls_det}) may be regularized and appropriate error indicators may be utilized to estimate an optimal value for $\epsilon$ or $r$.
\end{remark}

\subsection{ALS Algorithm and Computational Cost}
\label{sec:algrthm}
Algorithm~\ref{alg:prcss} summarizes the SR approximation process.

\begin{algorithm}[H]\label{alg:prcss}
\SetAlgoLined
$r = 0$\;
\While{$\gamma > \epsilon$}{
$r = r + 1$\;
Initialize $\bm{c}^r_k$, $k=1,\dots,d$, $\bm{u}_0^r$, and $s^r$\;
\While{ $\gamma$ decreases more than $\delta$ (See end of Section \ref{sec:als})}{
\For{$k = 1$ \KwTo $d$}{
Solve for $\bm{c}^l_{k}$ as elements of $\bm{z}$ using least squares problem (\ref{eq:ls})\;
Update ${s^l}$ and $\bm{c}^l_{k}$ using (\ref{eq:s_update})\;
}
Solve for $\bm{u}_0^l$ as columns of $\bm{Z}$ using least squares problem (\ref{eq:ls_det})\;
Update ${s^l}$ and $\bm{u}_0^l$ using (\ref{eq:u_tilde_update})\;
Generate $\hat{\bm{q}}(\bm\xi)$ using (\ref{eq:SR_vec})\;
Calculate $\gamma$ using (\ref{eq:conv});
}
}
\caption{Algorithm for an SR approximation}
\end{algorithm}

\vspace{.3cm}
As discussed in~\cite{beylkin}~and~\cite{doostan}, when $N\gg rP$, the total cost of generating and solving the least squares problems (\ref{eq:ls}) and (\ref{eq:ls_det}) for a full ALS sweep is $\mathcal{O}(d\cdot r^2\cdot P^2\cdot M\cdot N)$, which is linear in $d$. The total number of unknowns in the vector-valued SR (\ref{eq:SR_vec}) is $r\cdot P\cdot d+r\cdot M$, which is linear in $d$, assuming that the separation rank $r$ is independent of $d$. This therefore suggests a linear dependence of the number of samples $N$ on $d$, i.e.,
{
\begin{equation}\label{eq:costN}
N\sim\mathcal{O}(r\cdot P\cdot d+r\cdot M)
\end{equation}
}%
for a successful SR computation. Assuming this estimate for $N$, the total cost of a full ALS sweep grows quadratically in $d$. We highlight that, for situations where evaluating the QoI is significantly expensive, this cost is reasonable.

{ Due to the comparison between SR and PC expansions presented in this paper, the computation cost of the latter is discussed. For results discussed in Section~\ref{sec:rslts}, the PC expansion is computed using a least squares regression. In the case of a PC expansion, the number of required samples $N$ is given by~\cite{hampton15b} as
\begin{equation}\label{eq:pc_cst}
N \sim \mathcal{O} \left( \Lambda_{PC} \right),
\end{equation}

\noindent where $\Lambda_{PC}$ is the total number of PC terms
\begin{equation}\label{eq:lambdaPC}
\Lambda_{PC} = \frac{\left(P_{PC} + d\right)!}{\left(P_{PC}\right)!d!}.
\end{equation}
\noindent In (\ref{eq:lambdaPC}), $P_{PC}$ is the total order of the expansion. The methodology of using least squares determines the cost of creating a PC expansion, which produces (\ref{eq:pc_cst}) and (\ref{eq:lambdaPC}). 
\begin{remark}\label{rmrk:cst}
As seen in (\ref{eq:pc_cst}) and its reliance on (\ref{eq:lambdaPC}), the required number of samples $N$ for a PC expansion increases much faster with respect to dimension $d$ than the largely linear relationship of SR, i.e. $N\sim\mathcal{O}(r\cdot P\cdot d+r\cdot M)$.
\end{remark}}

\section{Solution Statistics and Sensitivity Analysis}\label{sec:sltnstt}

Once the coefficients $\{\bm{c}^l_i\}$ and $\{\bm u_0^l\}$ have been calculated, the statistics of $\bm{q}(\bm{\xi})$ may be approximated using the surrogate model $\hat{\bm{q}}(\bm{\xi})$. This can be done by sampling $\hat{\bm{q}}(\bm{\xi})$ using a Monte Carlo method or analytically from the coefficients $\{\bm{c}^l_i\}$ and the deterministic modes $\{\bm{u}_0^l\}$. Either of these methods may be applied to reduce the computation time of solution statistics when compared to traditional Monte Carlo methods. {Notice that the accuracy of the approximate statistics depends on the accuracy of $\hat{\bm{q}}(\bm{\xi})$, which may be first verified based, for instance, on validation experiments as illustrated in the examples of Section \ref{sec:rslts}.}

The analytical mean and second moment of an SR solution are derived in~\cite{doostan}. The mean for the $m$th entry of $\bm{q}(\bm{\xi})$ is
\begin{equation}\label{eq:anlmn}
E\left(\hat{{q}}_m\right) = \sum\limits_{l = 1}^r s^l \, u_{0,m}^l \prod\limits_{i = 1}^d c_{i,1}^l,
\end{equation}
%
\noindent {When considering the $m \times m'$ covariance matrix for $m = 1\ldots M$ and $m' = 1\ldots M$, each value $COV_{mm'} = E\left((\hat{{q}}_m - E(\hat{{q}}_m))(\hat{{q}}_{m'} - E(\hat{{q}}_{m'}))^{\text{T}}\right)$ is computed as
\begin{equation}\label{eq:anlstd}
 COV_{mm'}= \sum\limits_{l = 1}^r \sum\limits_{l' = 1}^r s^l \, u_{0,m}^l \, s^{l'} \, u_{0,m'}^{l'} \prod\limits_{i = 1}^d\left(\sum\limits_{p = 1}^P c_{i,p}^l c_{i,p}^{l'}\right) - E\left(\hat{{q}}_m\right)E\left(\hat{{q}}_{m'}\right).
\end{equation}
\noindent Using (\ref{eq:anlstd}), the variances of $\hat{q}$ are found along the diagonal, or when $m = m'$.}

For the low-order moments, these analytical methods are computationally more efficient than a traditional Monte Carlo sampling performed on $\hat{\bm{q}}(\bm{\xi})$. In general, higher order moments of $q_m$ may be estimated using a method similar to (\ref{eq:anlmn})~and~(\ref{eq:anlstd}), but a statistical sampling method of $\hat{q}_m$ may also be employed~\citep{doostan}. This sampling method may also be used to construct the PDF of $\bm{q}$. By evaluating the solution $\hat{\bm{q}}(\bm{\xi}_j)$ for large $N$, an estimate of the PDF of $\bm{q}$ is found without a large number of ODE solves. The results of this method can be seen in Section~\ref{sec:rslts}.

In addition to the statistics of $\bm{q}(\bm{\xi})$, a sensitivity analysis of the solution may be performed. The results of such an analysis determine the relative effect that a random input has on the the uncertainty of each QoI. \cite{sobol} discusses such an analysis and the difference between local and global sensitivities. Local sensitivities specify the derivative of the solution with respect to a stochastic input at a given realization of the inputs. Global sensitivities, however, consider the entire solution rather than a single solution realization~\citep{sobol,saltelli}.

The advantage of an SR approach lies in its separated structure and the relationship between computation costs and the dimension of $\bm\xi$. A larger dimension can be analyzed tractably when compared to methods that suffer from the curse of dimensionality. Taking advantage of a large dimension ensures a more complete view of the sensitivities, which leads to more informed decisions on system design and the selection of random inputs for operations. The surrogate approach of PC has been shown to reduce computation cost over a Monte Carlo based sensitivity analysis~\citep{sudret}, and it is expected that SR would further reduce these computation costs, when considering a large stochastic dimension. It should be noted that, due to its polynomial nature, SR can readily produce a local sensitivity based on derivatives with respect to the stochastic dimension. 

This paper, however, proposes determining the global \textit{sensitivity indices} from the SR solution using a Sobol approach,~\cite{saltellis}. Specifically, the presented method determines the variability of each component of the approximate QoI, $\hat{q}_m$, with respect to each direction $i$. Following \cite{saltellis}, the (total) sensitivity indices, $\left\{S_{i,m}\right\}^M_{m = 1}$, are given by 
\begin{equation}
\label{eq:sense}
S_{i,m} = \frac{V\left(E\left(\hat{q}_m|\xi_i\right)\right)}{V\left(\hat{q}_m\right)}=\frac{U_{i,m} - \left(E(\hat{q}_m)\right)^2}{V\left(\hat{q}_m\right)},
\end{equation}
where
\begin{equation}
U_i=\int \left(E(\hat{q}_m | \xi_i = \tilde{\xi}_i)\right)^2\rho_i(\tilde{\xi}_i)\text{d}\tilde{\xi}_i
\end{equation} 
and $\rho_i(\xi_i)$ is the marginal density function of $\xi_i$, here of standard Gaussian type. In practice, as detailed in \cite{saltellis}, $U_{i,m}$ in (\ref{eq:sense}) is estimated by MC sampling, 
\begin{equation}
\hat{U}_{i,m} = \frac{1}{N - 1}\sum_{j = 1}^N \hat{q}_m\left( \xi_{1,j}, \ldots, \xi_{d,j}\right)\hat{q}_m\left( \xi_{1,j}^{\prime},\ldots, \xi_{(i-1),j}^{\prime}, \xi_{i,j}, \xi_{(i+1),j}^{\prime},\ldots, \xi_{d,j}^{\prime}\right),
\end{equation}
where $\{\bm\xi_j\}_{j=1}^{N}$ and $\{\bm\xi_j^{\prime}\}_{j=1}^{N}$ are two sets of independent random samples of $\bm\xi$.

We highlight that, as compared to the standard formulation where the indices $S_{i,m}$ are computed via MC simulation using a large number of ODE solves, here the MC simulation is performed on the SR solutions $\hat{q}_m$. Therefore, the overall cost of determining such sensitivity information is negligible, once the SR is generated, and a large number of samples, $N$, may be generated for an accurate estimation of $\hat{U}_{i,m}$. 

%

%
\section{Numeric Results}\label{sec:rslts}

When considering Sections~\ref{sec:prbstp}~and~\ref{sec:SR}, test cases for this paper seek the estimation of the function $\bm{q}(\bm{\xi})$. As discussed in Section~\ref{sec:prbstp}, realizations of this function are the position and velocity state of a satellite at a considered time. {These states are propagated using an ODE integrator and are ultimately derived from an initial condition and Gaussian random variables as seen in Section~\ref{sec:apprx}}. Therefore, the final states $\{\bm{q}(\bm{\xi}_j)\}$ are treated as training samples and, along with associated random variables $\{\bm{\xi}_j\}$, are used to estimate the coefficients $\{\bm{c}_i^l\}$ and $\{{\bm{u}}_0^l\}$. These solutions are then used to create an approximate PDF by evaluating many more sets of samples of the inputs. {As we shall demonstrate, the QoI considered in the following test cases admit small separation ranks $r$, which range from $r=3$ to $r=6$ depending on the case, for the considered accuracies. Such small separation ranks lead to accurate estimation of statistics of the QoIs, its PDFs, and sensitivities with respect to the input variables, using relatively small numbers of samples of QoI. For scenarios where the separation rank is not small, one shall not anticipate similar accuracies as in these experiments.}

Each test compares the performance of SR with that of a PC result that is converged with respect to STD. PC is chosen as a reference solution due to its proven nature of converging efficiently and accurately~\citep{jones}. In the case of high stochastic dimension, PC is still used, albeit with high computation cost due to the \textit{curse of dimensionality}. All test cases incorporate a $50\times50$ spherical harmonics model of the Earth's gravity perturbations, as determined by GRACE GGM02C gravity model~\citep{tapley}, as well as atmospheric drag perturbations based on the exponential cannonball model presented in~\cite{vallado}. All cases are propagated for 36 hours using a Dormand-Prince 5(4) integrator with a tolerance of $10^{-13}$.

The first test case considers STDs of 1 km and 1 m/sec in the initial position and velocity, respectively, generating a stochastic dimension of $d = 6$. The second case is similar to the first, with the exception that 14 random inputs (to bring the total to $d = 20$) are added in the form of uncertainty in the low degree Stokes coefficients, drag parameters and the gravitational parameter. The Stokes coefficients used as random inputs are included in the \nameref{sec:appndx}. Initial values and uncertainties are taken from~\cite{tapley}. Parameters for the \textit{a priori} state PDF for Test Case 2 can be seen in Table~\ref{tbl:std_20}. Test Case 1 uses the first six random inputs in the form of ECI coordinates and velocity. In both test cases, ECI position and velocity as a function of the random inputs are estimated using the SR surrogate for a vector-valued function.

For the third test, a problem presented in~\cite{horwood} is analyzed using SR in place of the original examination with a GMM. In this case, ECI coordinates are replaced with equinoctial elements. This coordinate system is composed of the semimajor axis and five additional elements. The equations for these additional elements as a function of the Keplerian elements are included in the \nameref{sec:appndx}. Each sample in equinoctial elements is transformed into ECI coordinates and propagated for 36 hrs. The propagated ECI state is then transformed back to equinoctial elements to be used as a training sample. In this case, all six equinoctial elements are estimated using the SR estimation process for a vector-valued function. All random inputs and their relevant uncertainties are presented in Table~\ref{tbl:std_equi}.

\begin{table}[h]
\begin{center}
\caption{Random inputs and associated STDs for the first two test cases}
\begin{tabular}{lll}\toprule &Mean&STD\\ \midrule $x$ (km)&757.700&1.0\\$y$ (km)&5222.607&1.0\\ $z$ (km)&4851.800&1.0\\ $\dot{x}$ (m/s)&2213.210&1.0\\$\dot{y}$ (m/s)&4678.340&1.0\\$\dot{z}$ (m/s)&-5371.300&1.0\\$\mu$ (km$^3/$s$^2$)&$3.986\times10^5$&$10^{-3}$\\$C_D$&2.0&$0.398$\\$A/m$ (m$^2/$kg)&0.01&$1.7\times10^{-3}$\\\bottomrule
\end{tabular}\label{tbl:std_20}
\end{center}
\end{table}

\begin{table}[h]
\begin{center}
\caption{Random inputs and associated STDs for Test Case 3}
\begin{tabular}{lll}\toprule &Mean&STD\\ \midrule $a$ (km)&6980.0&20.0\\h&0.0&$10^{-3}$\\ $k_e$&0.0&$10^{-3}$\\$p_e$&0.0&$10^{-3}$\\$q_e$&0.0&$10^{-3}$\\$\lambda_\mathcal{M}$ (rad)&0.0&$10^{-2}\frac{\pi}{180}$\\\bottomrule
\end{tabular}\label{tbl:std_equi}
\end{center}
\end{table}

In all cases, canonical units are utilized for estimation and plotting. Therefore, all units of distance have been normalized with respect to the Earth's mean radius ($r_{\oplus} = 6371$ km), resulting in distance units DU, while seconds have been normalized with the time unit TU, where

\begin{equation}
\text{TU} = \sqrt{r_{\oplus}^3/\mu},
\end{equation}

\noindent and $\mu$ is the gravitational parameter. The result of using canonical units is a data set that is mostly of the same magnitude with respect to quantities of interest and therefore more numerically stable. 

\subsection{Test Case 1}

\begin{figure}[h]
 \centerline{\includegraphics[width = \textwidth]{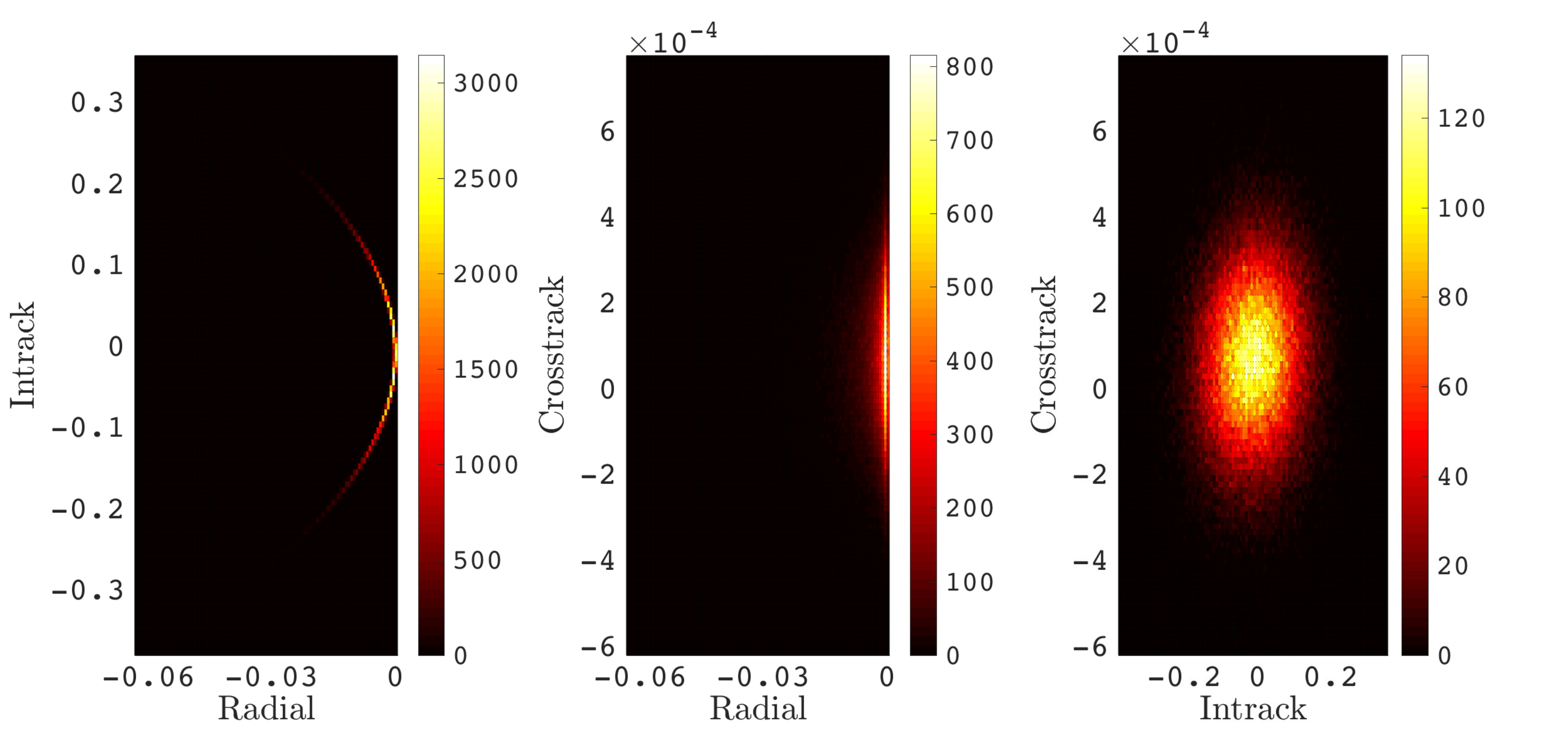}}
\caption{{MC results for Test Case 1 plotted in RIC coordinates. Note that the colorbars illustrate object count.}}
 \label{fig:RIC_d6}
\end{figure}

For an initial example, a case of a satellite in low Earth orbit is examined. {The initial conditions for position and velocity in Table~\ref{tbl:std_20} are used as the initial mean solution, with the respective uncertainties being applied with the methodology described in Section~\ref{sec:apprx}.} Figure~(\ref{fig:RIC_d6}) depicts the distribution of Radial-Intrack-Crosstrack (RIC) coordinates after 36 hours when considering 100,000 MC samples.All RIC plots (MC- and SR-based) are generated using the same MC propagated position and velocity sample, pulled from the training data set, as the center of the coordinate frame. This sample is unique for each test case, and the particular value was randomly chosen as it puts the origin of the frame at a possible realization of the posterior PDF. As depicted, the \textit{a posteriori} position distribution is non-Gaussian. This is most evident in the Radial-Intrack plot, where a large amount of skewness is evident. Hence, any Gaussian assumptions on the posterior are invalid when attempting to accurately estimate the PDF. Therefore, this case requires a higher-fidelity method.

\begin{figure}[h] 
 \centerline{\includegraphics[width = \textwidth]{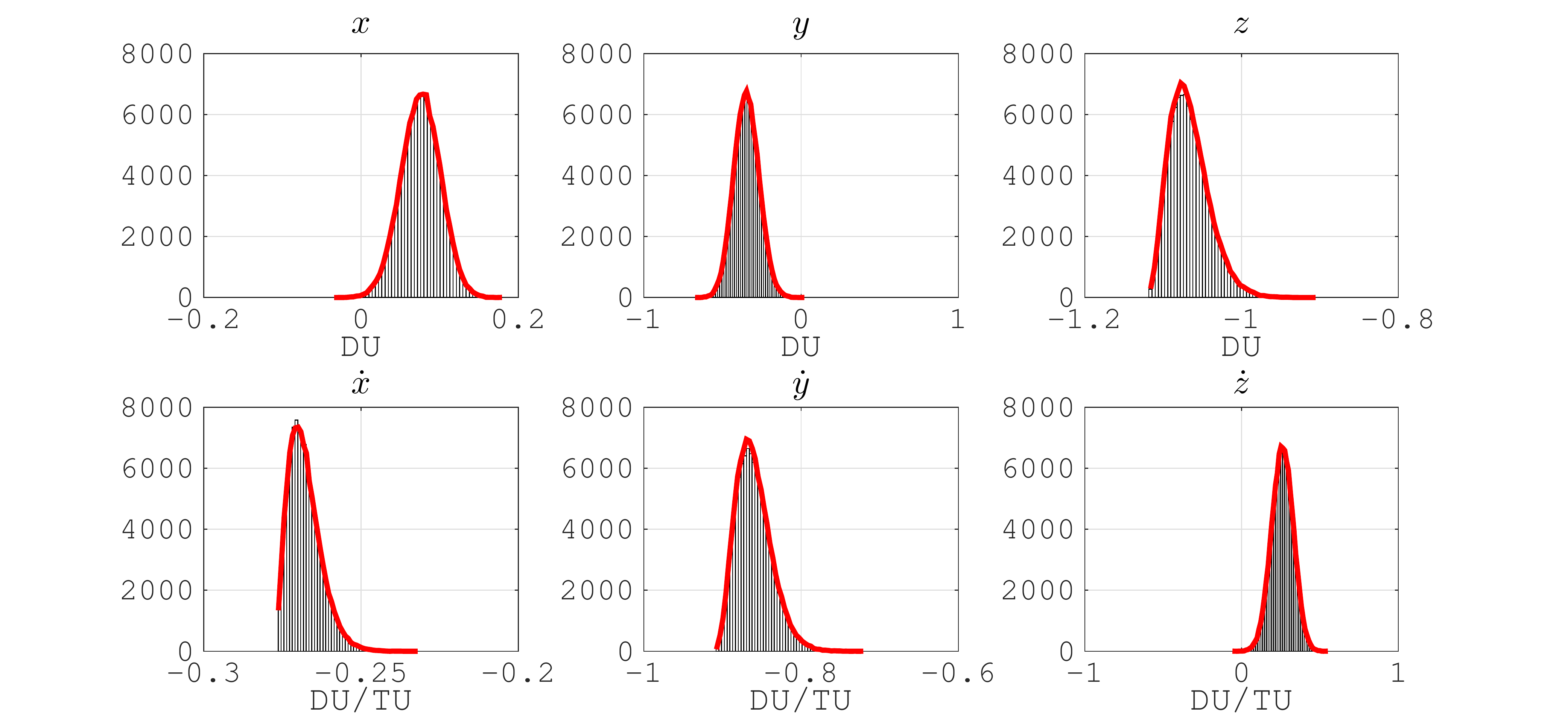}}
\caption{Histograms of SR results for quantities of interest in Test Case 1.}
 \label{fig:hist_d6}
\end{figure}

{Using 350 training samples, $r = 5$, $P = 4$ and $\delta = 10^{-7}$}, an SR surrogate is created for approximating the ECI state of the satellite. {In order to visualize the distributions of the test case and characterize the accuracy of the SR-based PDF, MC runs of $100,000$ samples are created for Test Case 1 and $100,000$ evaluations of an SR are created using independent sets of samples of the inputs and the appropriate uncertainties. This approach is then used in all test cases, producing Figs.~(\ref{fig:hist_d6}),~(\ref{fig:hist_d20})~and~(\ref{fig:hist_equi}) in which these density functions are compared.} The histograms of the MC results are displayed as a solid line (derived from interpolating the centers of the bins) plotted over the histograms of the SR-generated samples. This method allows for qualitative assessments of SR's ability to capture the third and higher moments. Each test case is also analyzed quantitatively. These analyses consist of relative residuals calculated for the mean and STD of QoIs. The relative residual provides information for knowing the digits of accuracy in a solution and is calculated by

\begin{equation}
\epsilon_{rel} = \left\vert\frac{\lambda - \hat{\lambda}}{\lambda}\right\vert,
\end{equation}

\noindent where $\epsilon_{rel}$ is the relative residual, $\lambda$ is a reference value, and $\hat{\lambda}$ is an estimate of this reference. {Additionally, the accuracy of each surrogate is estimated by evaluating absolute residuals of a small set of random samples, which are not used in the training of the surrogate. This process computes the difference between the deterministic solution using the block-box propagator and the surrogate-based solution, with each method using the same random inputs. The RMS of the difference is taken over the number of validation samples and provided as a quantitative measure for determining the {\it goodness} of the surrogate and solution convergence. Such a strategy is referred to as {\it cross-validation} (or {\it validation} for brevity) and may be extended to multiple constructions of the surrogate and residual evaluation on independent samples of the QoI, see, e.g., \cite[Chapter 7]{friedman2001elements}. { Validation results are included for each test case in Tables~\ref{tbl:val_1},~\ref{tbl:val_2},~\ref{tbl:valRIC_2}~and~\ref{tbl:val_3}, which include the RMS of the difference taken over the number of validation samples as well as the RMS of the MC-based samples for comparison.}}

With (\ref{eq:anlmn})~and~(\ref{eq:anlstd}), means and STDs for each random input of the satellite's state are compared to those of a converged PC solution. The results can be seen in Table~\ref{tbl:stats_1}. Since the reference solution is converged with respect to STD, relative residuals for the third and fourth moments are not provided. {Table~\ref{tbl:val_1} includes the validation results for Test Case 1. The table presents RMS values of the absolute residuals for 70 independent samples of random inputs, as well as the RMS values of the MC-based validation samples. {By comparing the RMS of the residuals with the RMS of the MC-based validation samples, we observe the accuracy of the SR in predicting the states at samples of inputs not included in the training samples.}} 

\begin{figure}[h] 
 \centerline{\includegraphics[width = \textwidth]{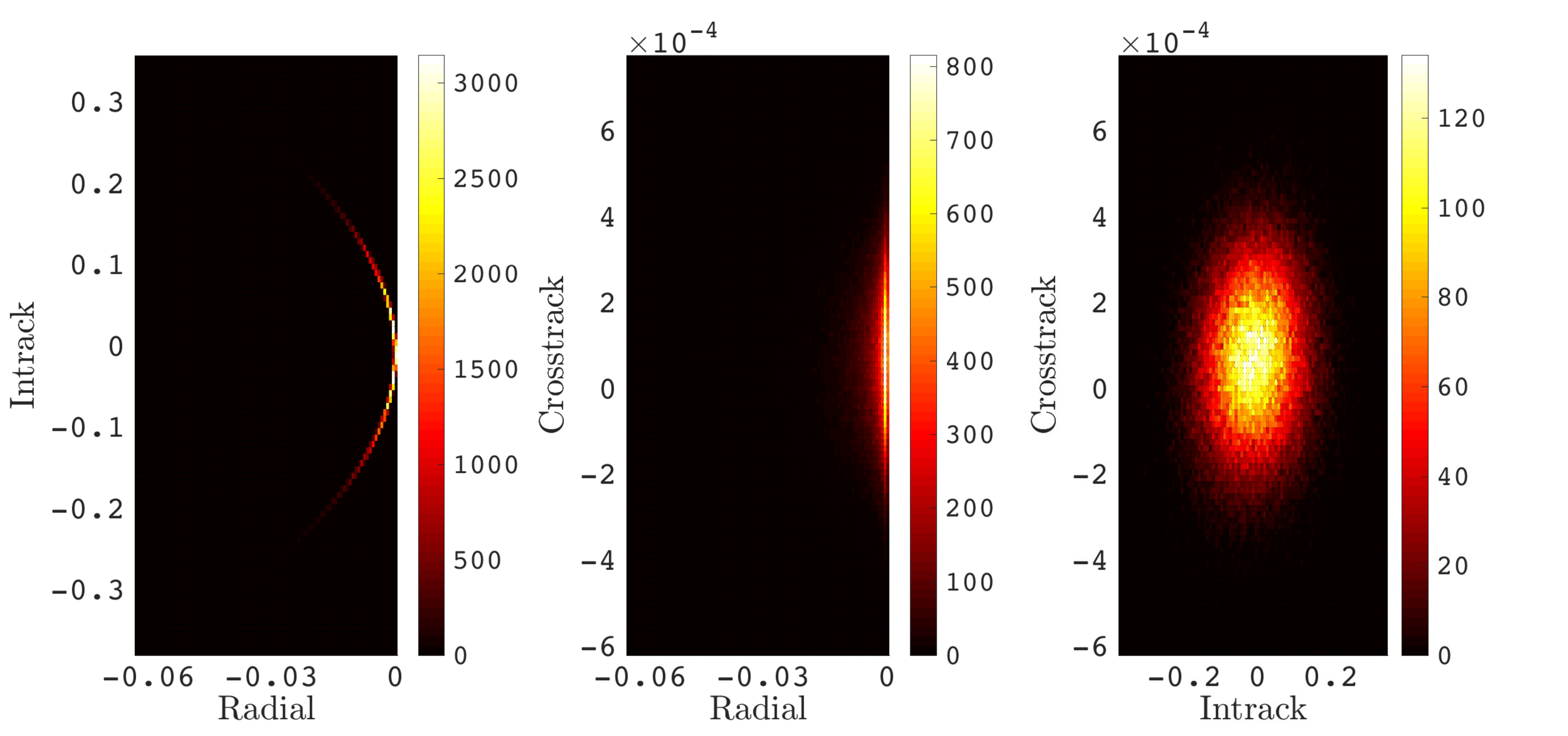}}
\caption{{SR results for Test Case 1 plotted in the RIC coordinates. Note that the colorbars illustrate object count.}}
 \label{fig:RICd6SR}
\end{figure}

Qualitatively, it can be seen in Figure~\ref{fig:hist_d6} that the SR solution captures the non-Gaussian distributions well. In the case of $\dot{x}$, the distribution is highly non-Gaussian. The histogram of the SR solution follows the MC distribution well, capturing the overall skewness and kurtosis. { The results of the SR evaluation are also transformed to the RIC frame and plotted in Figure~\ref{fig:RICd6SR}.} In addition to these qualitative fits, three digits of precision or more are captured in the first and second moments when comparing the converged PC solution to that of SR. 
\begin{table}[h]
\begin{center}
\caption{Agreement between SR- and PC-based mean and STD for Test Case 1}
\begin{tabular}{lllll}\toprule  
	 &Ref. Mean       &Ref. STD	    &Rel. Mean&Rel. STD \\ \midrule 
$x$	 &0.075765 (DU)    &0.02498 (DU)      &9.8e-05  &2.7e-04\\ 
$y$	 &-0.34683 (DU)    &0.07986 (DU)      &6.5e-05  &1.8e-04\\ 
$z$	 &-1.06680 (DU)    &0.02534 (DU)      &3.4e-06  &1.5e-04\\ 
$\dot{x}$&-0.26786 (DU/TU) &5.13e-03 (DU/TU)  &2.0e-05  &1.0e-03\\ 
$\dot{y}$&-0.859985 (DU/TU)&0.02238  (DU/TU)  &4.3e-06  &1.8e-04\\
$\dot{z}$&0.2603 (DU/TU)   &0.06937  (DU/TU)  &1.3e-04  &3.0e-04\\ \bottomrule
 \end{tabular}\label{tbl:stats_1}
\end{center}
\end{table}
\begin{table}[h]
\begin{center}
\caption{{Residual RMS of 70 SR- and MC-based validation samples for Test Case 1}}
\begin{tabular}{llll}\toprule  &MC Sample RMS&Residual RMS&Units\\ \midrule 
$x$      &0.07679&1.0e-04&(DU)\\ 
$y$      &0.3705 &3.9e-04&(DU)\\ 
$z$      &1.0620 &2.9e-05&(DU)\\ 
$\dot{x}$&0.2685 &1.2e-04&(DU/TU)\\ 
$\dot{y}$&0.85600&4.7e-05&(DU/TU)\\
$\dot{z}$&0.2823 &3.0e-04&(DU/TU)\\ \bottomrule
 \end{tabular}\label{tbl:val_1}
\end{center}
\end{table}
%
%
\subsection{Test Case 2}

For Test Case 2, the previous initial conditions are kept but the stochastic dimension is expanded to include all elements from Table~\ref{tbl:std_20}, in addition to the Stokes coefficients from Table~\ref{tbl:stokes}. Theoretically, an estimation method that suffers from the \textit{curse of dimensionality} would require significantly more samples than a method, such as SR, that increases the number of samples linearly with respect to stochastic dimension. {This cost comparison is elaborated upon in Section~\ref{sec:algrthm} and its associated Remark~\ref{rmrk:cst}.}

\begin{figure}[h] 
 \centerline{\includegraphics[width = \textwidth]{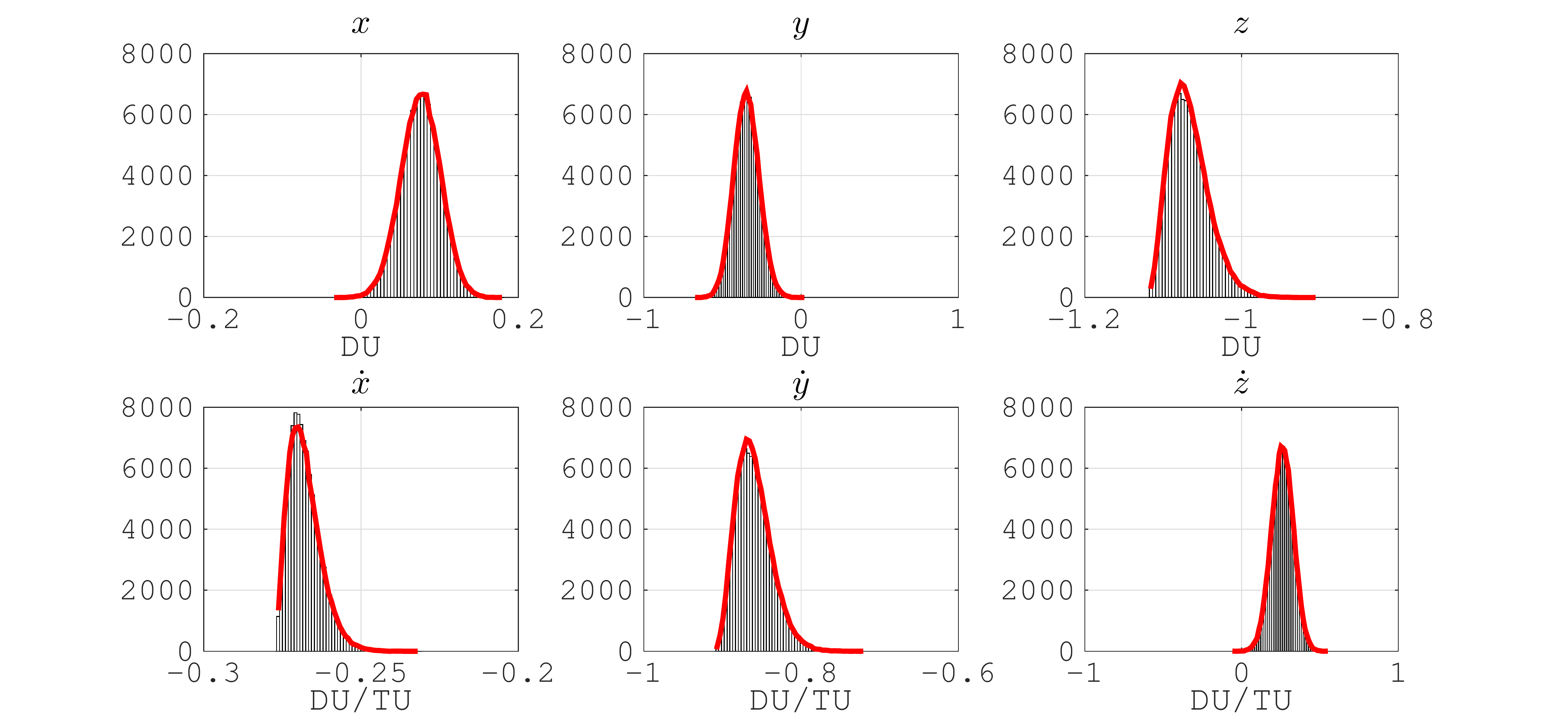}}
\caption{Histograms of SR results for quantities of interest in Test Case 2.}
 \label{fig:hist_d20}
\end{figure}

{Using $N = 750$ training samples, $P = 4$ and $\delta = 10^{-7}$, an SR surrogate is created for estimating the ECI state of the satellite. In this case, results are included for surrogates of rank 3, 4 and 5 with each using the same 750 training samples. By presenting results for these three choices of $r$, a clearer picture of the convergence of an SR surrogate can be seen. Figure~\ref{fig:RIC_d20} presents the 100,000 MC realizations in the RIC frame, which resembles that of Figure~\ref{fig:RIC_d6}. This figure can then be compared to the results presented in Figures~\ref{fig:cmpRI},~\ref{fig:cmpRC},~and~\ref{fig:cmpIC}. Each of these figures illustrates realizations generated by the SR surrogates of ranks 3, 4 and 5. Convergence to the MC distributions can been seen as the ranks increase, with $r = 5$ providing a good qualitative fit.

Table~\ref{tbl:val_2} includes the quantitative validation results for Test Case 2. The table presents RMS values of the  residuals for 150 independent random input vectors as evaluated by either a rank 3, 4 or 5 surrogate, as well as RMS values of the MC-based validation samples. Although no improvement in accuracy may be readily apparent in the ECI frame, when the error is transformed to the RIC frame as displayed in Table~\ref{tbl:valRIC_2}, the estimate of the crosstrack state improves with each additional rank. By comparing the validation results of the MC-based RMS with the residual RMS, the sample RMS in the crosstrack direction indicates a smaller mean compared to the radial and intrack directions. In this case, the $r = 5$ case is the only solution to yield a residual RMS smaller than the sample RMS. As illustrated in Figures~\ref{fig:cmpRC}~and~\ref{fig:cmpIC}, the crosstrack QoI lies within bounds on the order of $10^{-4}$. As seen in Table~\ref{tbl:valRIC_2}, the crosstrack error magnitude for ranks 3 and 4 indicates that one digit of accuracy has not been achieved. The improvement of adding a fifth rank, however, is enough for the crosstrack accuracy of the rank 5 solution to be on the order of $10^{-5}$ and therefore accurate enough to capture the crosstrack distribution.

As this discussed investigation utilizes the \textit{a priori} knowledge found in Figure~\ref{fig:RIC_d20}, an operational implementation concerned with crosstrack accuracy would require a different approach. Table~\ref{tbl:valRICstd_2} contains STD values of Test Case 2 approximated by the previously discussed SR solutions and transformed into the RIC frame. By comparing these values to those of Table~\ref{tbl:valRIC_2}, it can be seen that the STDs of the crosstrack direction for the rank 3 and 4 surrogates are smaller than each correspoding residual RMS by an order of magnitude or more. The rank 5 approximation of crosstrack STD, however, is larger than the respective residual RMS. Although this is not proof of solution convergence to STD, it does explain the inability to capture information from the PDFs using the rank 3 and 4 solutions. With a residual RMS larger than the STD, the approximation of the crosstrack direction variance is largely due to the accuracy of the solution and not the actual PDF. It should be noted that although the approximate value of STD for the crosstrack direction, as seen in Table~\ref{tbl:valRICstd_2}, increases by an order of magnitude with each increase in rank, a solution of rank 6 yields a crosstrack STD of 1.6e-04 DU. This value compares well with the rank 5 approximation.\\
\begin{figure}[h] 
 \centerline{\includegraphics[width = \textwidth]{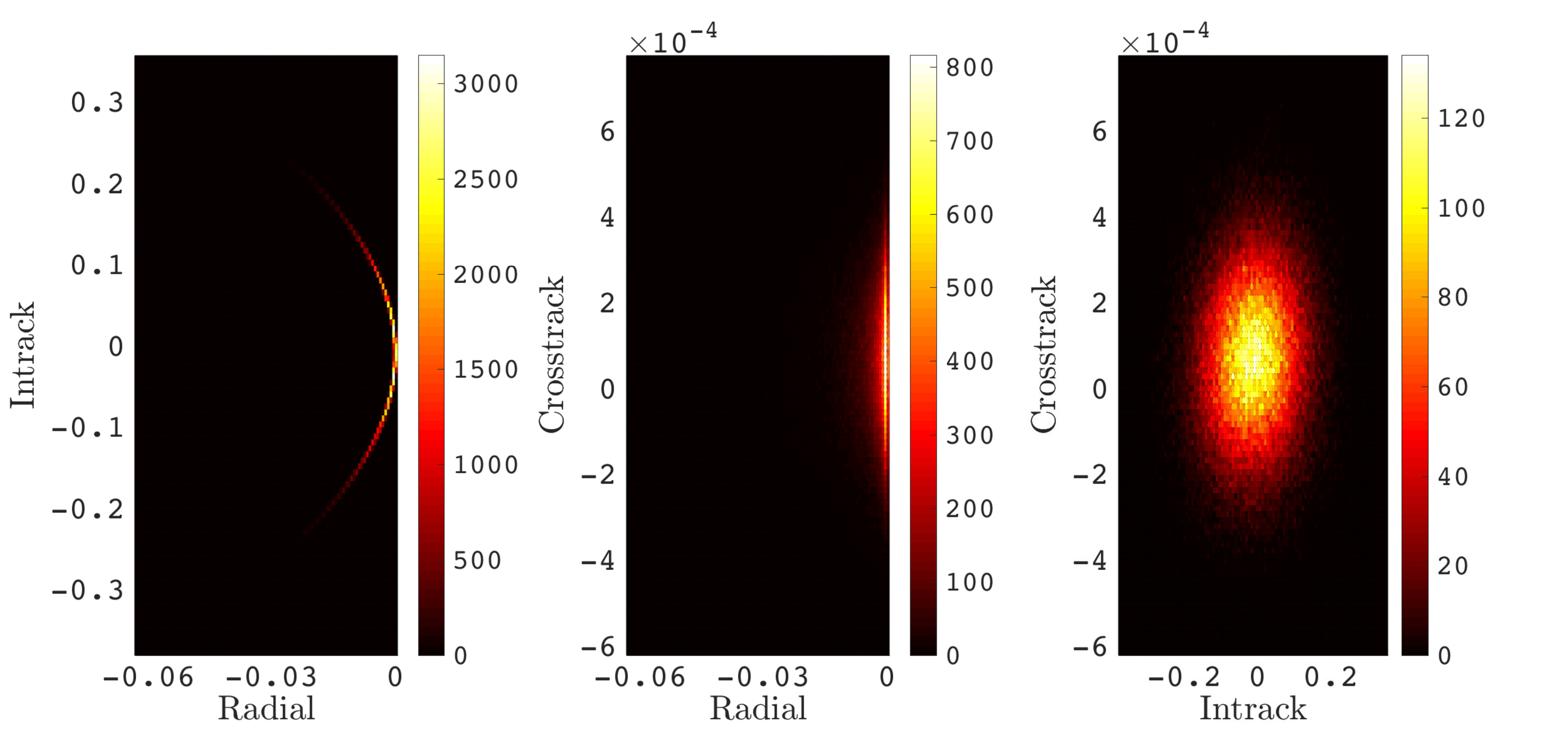}}
\caption{{MC results for Test Case 2 plotted in RIC coordinates. Note that the colorbars illustrate object count.}}
 \label{fig:RIC_d20}
\end{figure}
\begin{figure}[h] 
 \centerline{\includegraphics[width = \textwidth]{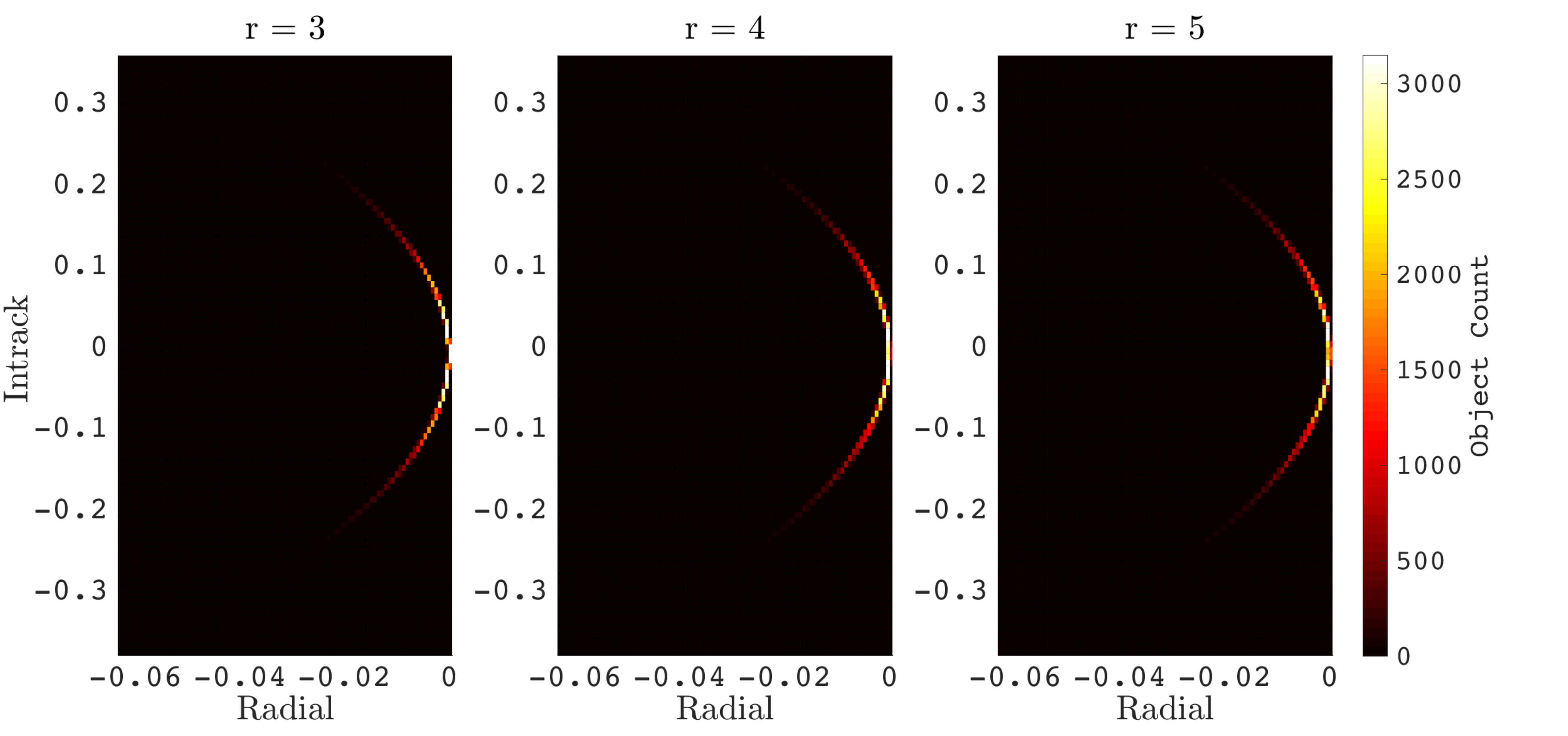}}
\caption{{SR results for Test Case 2 plotted as radial and intrack.}}
 \label{fig:cmpRI}
\end{figure}
\begin{figure}[h] 
 \centerline{\includegraphics[width = \textwidth]{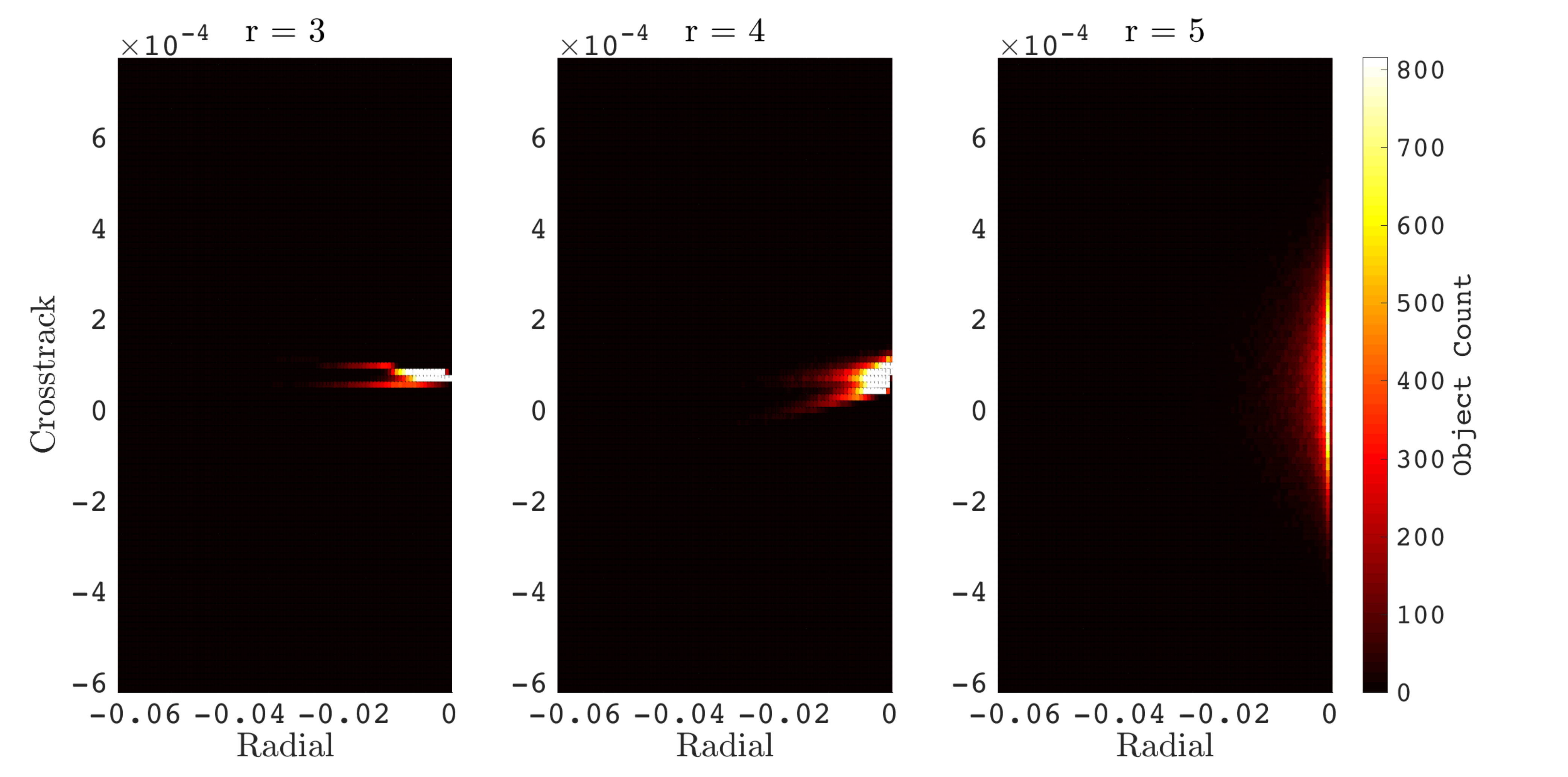}}
\caption{{SR results for Test Case 2 plotted as radial and crosstrack.}}
 \label{fig:cmpRC}
\end{figure}
\begin{figure}[h] 
 \centerline{\includegraphics[width = \textwidth]{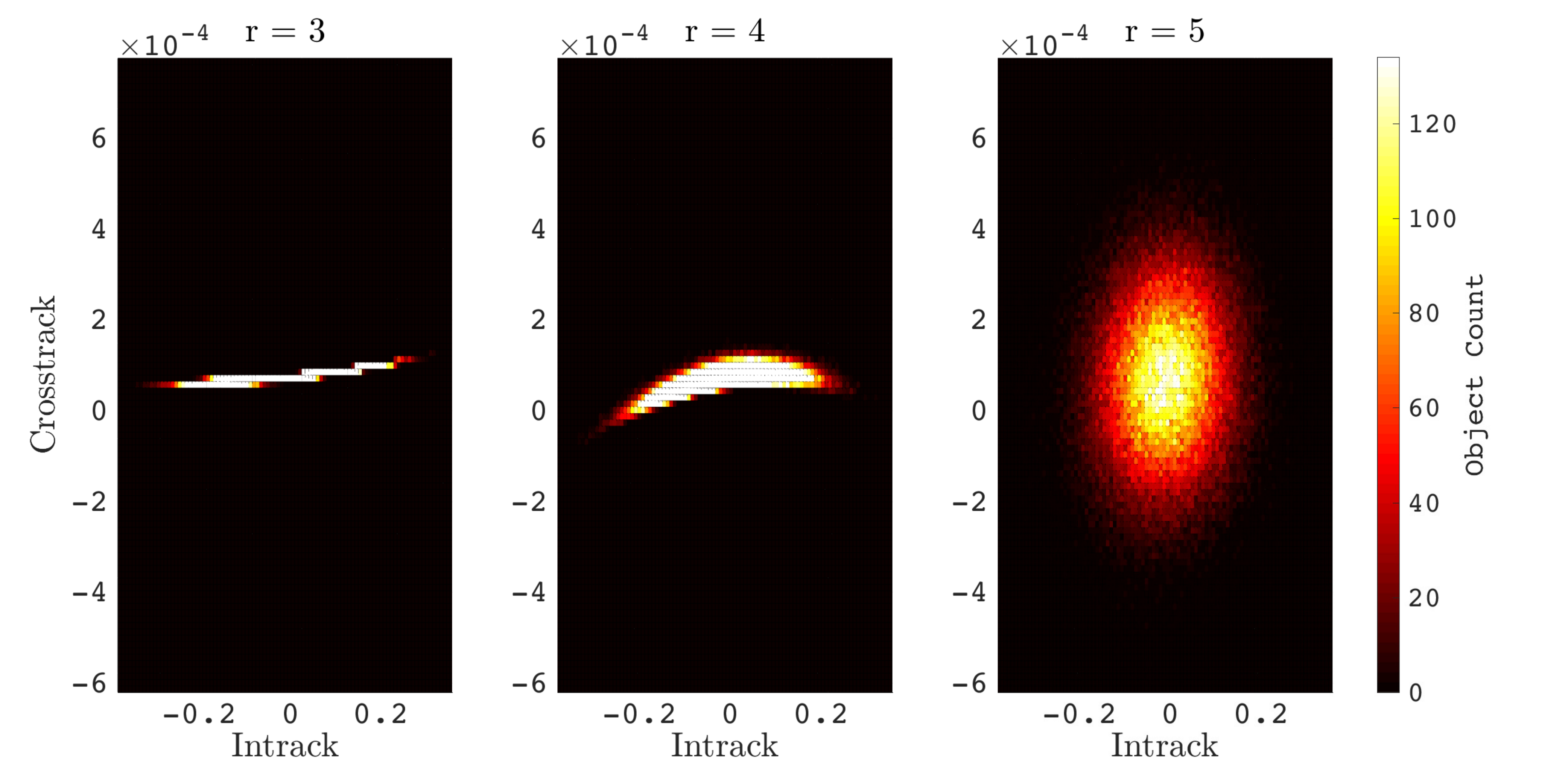}}
\caption{{SR results for Test Case 2 plotted as intrack and crosstrack.}}
 \label{fig:cmpIC}
\end{figure}
\begin{table}[h]
\begin{center}
\caption{{Residual RMS of 150 SR- and MC-based validation samples for Test Case 2}}
\begin{tabular}{llllll}\cmidrule(lr){3-5}  &&\multicolumn{3}{c}{Residual RMS}\\\midrule&MC Sample RMS&$r=3$&$r=4$&$r=5$&Units\\ \midrule 
$x$      &0.0790 &1.7e-04&1.6e-04&1.0e-04& (DU)\\ 
$y$      &0.3594 &3.8e-04&2.7e-04&2.7e-04& (DU)\\ 
$z$      &1.0658 &1.1e-04&1.9e-04&1.7e-04& (DU)\\ 
$\dot{x}$&0.2680 &1.2e-04&1.3e-04&1.1e-04& (DU/TU)\\ 
$\dot{y}$&0.8592 &1.3e-04&1.6e-04&1.4e-04& (DU/TU)\\
$\dot{z}$&0.2724 &3.2e-04&2.2e-04&2.2e-04& (DU/TU)\\ \bottomrule
 \end{tabular}\label{tbl:val_2}
\end{center}
\end{table}
%
%
\begin{table}[h]
\begin{center}
\caption{{Residual RMS of 150 SR- and MC-based validation samples in RIC frame for Test Case 2}}
\begin{tabular}{llllll}\cmidrule(lr){3-5}  &&\multicolumn{3}{c}{Residual RMS}\\\midrule&MC Sample RMS&$r=3$&$r=4$&$r=5$&Units\\ \midrule 
Radial      &1.1239&1.9e-04&2.5e-04&2.3e-04& (DU)\\ 
Intrack     &0.0896&3.6e-04&2.3e-04&2.2e-04& (DU) \\ 
Crosstrack  &1.8e-04&1.3e-04&1.3e-04&6.5e-05& (DU)\\  \bottomrule
 \end{tabular}\label{tbl:valRIC_2}
\end{center}
\end{table}
\begin{table}[h]
\begin{center}
\caption{{STD estimates in RIC frame for Test Case 2}}
\begin{tabular}{lllll}\cmidrule(lr){2-4}  &\multicolumn{3}{c}{STD}\\\midrule&$r=3$&$r=4$&$r=5$&Units\\ \midrule 
Radial      &4.8e-03&4.8e-03&4.8e-03& (DU)\\ 
Intrack     &8.7e-02&8.7e-02&8.7e-02& (DU)\\ 
Crosstrack  &8.3e-06&2.7e-05&1.5e-04& (DU)\\  \bottomrule
 \end{tabular}\label{tbl:valRICstd_2}
\end{center}
\end{table}
As $r = 5$ provides the best fit, further results are calculated with a surrogate corresponding to that rank.} Figure~\ref{fig:hist_d20} shows the SR result as a histogram plotted alongside a 100,000 MC result. Once again, it can be seen that the SR solution captures the non-Gaussian distributions well. The skewness and tails of the non-Gaussian distributions are represented in both the MC and SR PDFs. In addition to these qualitative fits, Table~\ref{tbl:stats_2} shows that three digits of precision or more are captured in the first and second moments, when compared to the PC result that has been converged with respect to STD. 
%
%
\begin{table}[h]
\begin{center}
\caption{Agreement between SR- and PC-based mean and STD for Test Case 2}
\begin{tabular}{lllll}\toprule  &Ref. Mean&Ref. STD&Rel. Mean&Rel. STD \\ 
\midrule 
$x$	 &0.075765  (DU)    &0.024985  (DU)     &1.4e-05  &4.0e-05\\ 
$y$	 &-0.346831 (DU)    &0.079863  (DU)     &4.6e-06  &3.0e-05\\ 
$z$	 &-1.066805 (DU)    &0.025340  (DU)     &2.9e-07  &3.2e-05\\ 
$\dot{x}$&-0.26786  (DU/TU) &5.135e-03 (DU/TU)  &1.8e-05  &5.8e-04\\ 
$\dot{y}$&-0.859985 (DU/TU) &0.02238   (DU/TU)  &1.5e-06  &3.2e-04\\
$\dot{z}$&0.260341  (DU/TU) &0.069376  (DU/TU)  &8.8e-04  &3.6e-05\\ \bottomrule
 \end{tabular}\label{tbl:stats_2}
\end{center}
\end{table}
%

%
%

\begin{figure}[h] 
 \centerline{\includegraphics[width = \textwidth]{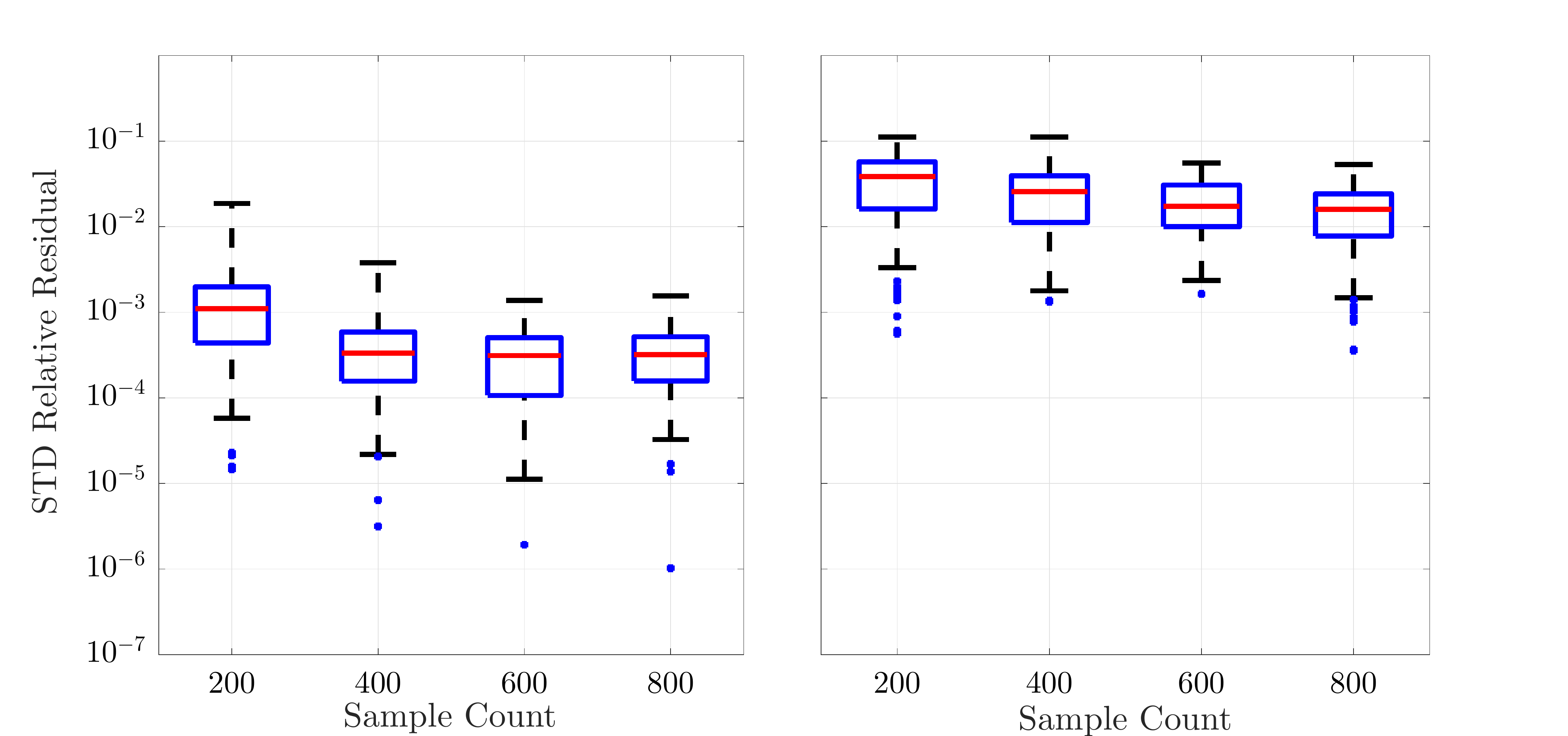}}
\caption{STD relative residuals. SR on left and MC on right}
 \label{fig:cnv}
\end{figure}

Figure~\ref{fig:cnv} illustrates the convergence of SR as a function of $N$ and compares it to that of MC. The figure shows the relative errors of the estimated STD for the $x$-position of 100 independent SR solutions. For each chosen $N$, 100 independent calculations of an SR surrogate and subsequent STD estimation are performed. This is done using the Test Case 2 initial conditions {with fixed SR parameters excluding the number of training samples.} The ordinate axes of the plots are logarithmic, the middle lines of the boxes are the medians, and the top edge of the boxes are the 75$^\text{th}$ percentiles (upper quartile) with the bottom edges being the $25^\text{th}$ (lower quartile). The upper and lower whiskers cover 1.5 times the interquartile range for the upper and lower quartiles, respectively, which is 99.3\% of the data if it was normally distributed. The remaining outliers are marked as blue dots. Here we use (\ref{eq:anlstd}) to approximate $\sigma$ for the SR solutions and sampling based method for the MC results. The fast convergence of SR estimates of STD (as a function of $N$) can be seen in the left plot of Fig.~\ref{fig:cnv}, and, when compared to the right plot, the relatively slow convergence of MC can also be seen.

\begin{figure}[h] 
 \centerline{\includegraphics[width = \textwidth]{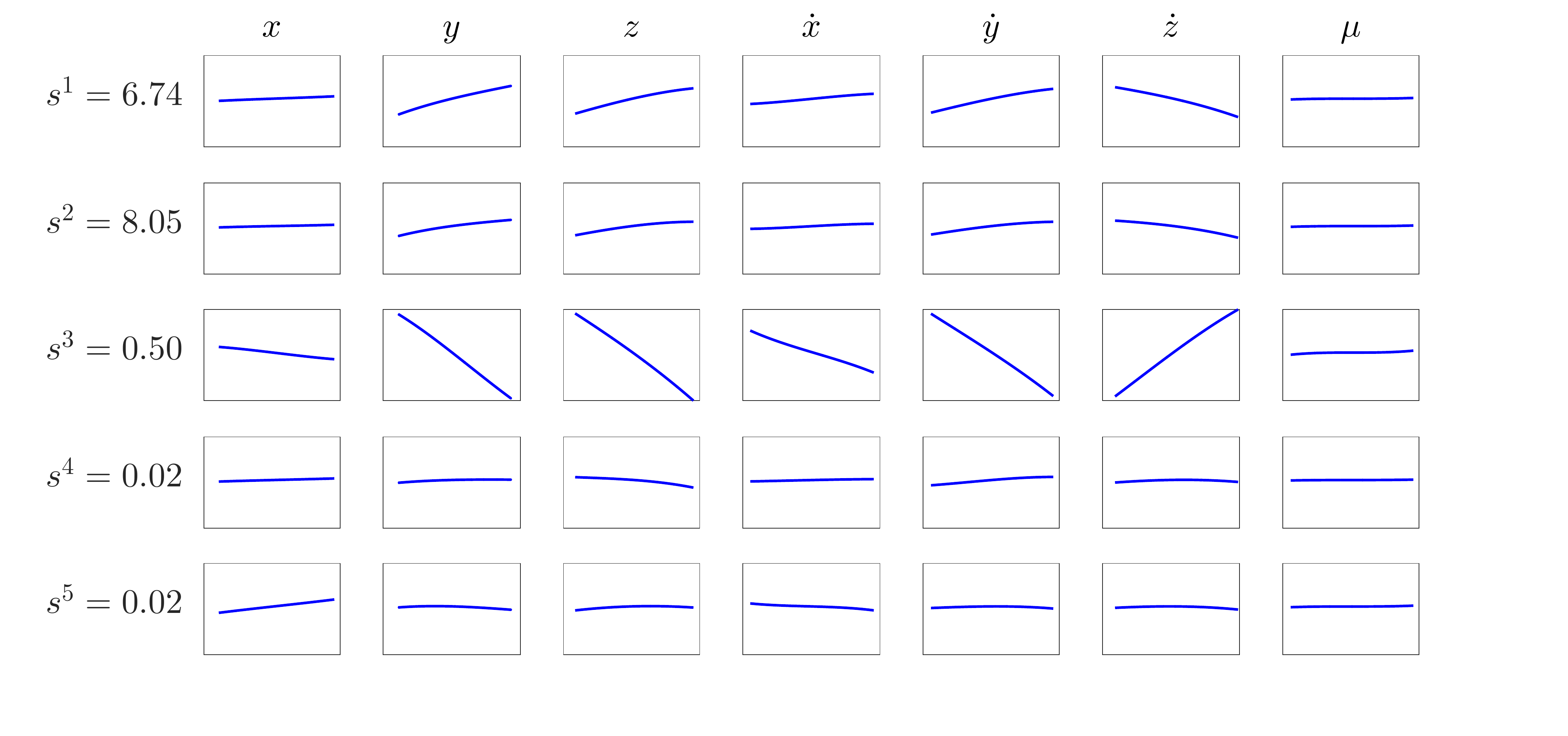}}
\caption{{Plot of the absolute values of the univariate factors $u_i^l(\xi_i)$ for Test Case 2.}}
 \label{fig:u_d20}
\end{figure}

A sensitivity analysis is applied to the results of Test Case 2. {Reference values found using a PC expansion are included in Tables~\ref{tbl:snsPos2}~and~\ref{tbl:snsVel2} along with absolute residuals found when approximating the indices with $10^6$ SR realizations. The PC results, which are generated such that they provide six digits of precision, utilize an analytic method for calculating sensitivity indices~\citep{sudret}, and serve as a baseline for assessing SR accuracy. Therefore, some values are stated as $\sim 0$, as the calculated indices are less than or equal to $10^{-6}$.}  { Each of these values represents, at most, $10^{-8}$ percent of the total variance contribution to a QoI due to the fact that total Sobol indices sum to one or greater~\citep{rsmithUQ}.}  {The indices in Tables~\ref{tbl:snsPos2}~and~\ref{tbl:snsVel2} show that the ECI random inputs contribute variability that is large enough to be quantified, with the $x$-position having the smallest contribution out of the six. In addition to these six random dimensions, the sensitivity index of $\mu$ is included. Using the indices as a guide, it can be concluded that uncertainty in $\mu$ produces little variability in the final solution for this test case. Figure~\ref{fig:u_d20} shows the absolute value of univariate functions $\{u_i^l\}$ for the random inputs provided in Tables~\ref{tbl:snsPos2}~and~\ref{tbl:snsVel2}. Constructed using $\{\bm{c}^l_i\}$, the appropriate polynomial bases and a set of samples of the inputs, the figure illustrates the variability of each element. The behavior of each $u_i^l$ is represented well by respective sensitivity indices. For example, the univariate functions of $x$ and $\mu$ exhibit less variability than the other shown random inputs. The results in Tables~\ref{tbl:snsPos2}~and~\ref{tbl:snsVel2} quantitatively reflect this, with $x$ and $\mu$ having low index values. In Fig.~\ref{fig:u_d20}, the values of $\{u_i^l\}$ for the other 13 random inputs are omitted due to the lack of variability. The sensitivity indices for the Stokes coefficients, $C_D$ and $A/M$ are all negligible (smaller than $10^{-6}$) and are omitted for brevity. The low sensitivity index values for these random inputs, including $\mu$, agree with intuition due to the physics of the high-altitude orbit. Gravitational perturbations as well as atmospheric drag effects are not significant in relation to position uncertainty at an altitude of around 790 km.}

\begin{table}[h]
\begin{center}
\caption{{Sensitivity indices $S_{i,m}$ and residuals of position QoIs for Test Case 2}}
\begin{tabular}{*{7}{c}}
\cmidrule(lr){2-7}
& \multicolumn{6}{c}{Quantities of Interest} \\ 
\midrule Random & \multicolumn{2}{c}{$x$}& \multicolumn{2}{c}{$y$}& \multicolumn{2}{c}{$z$} \\ 
\cmidrule(lr){2-7}
Inputs&PCE&Resid.&PCE&Resid.&PCE&Resid.\\\midrule  
$x$&         5.28e-03&8e-05&6.10e-03&3e-05&  6.43e-03&3e-05\\
$y$&         0.280   &9e-04&0.279   &1e-03&     0.285&1e-05\\
$z$&         0.235   &1e-03&0.236   &1e-03&     0.238&8e-04\\
$\dot{x}$&   0.0437  &4-e04&0.0423  &1-e04&     0.0440&2-e04\\
$\dot{y}$&   0.192   &1-e04&0.193   &2-e04&     0.201&6-e04\\
$\dot{z}$&   0.244   &2-e04&0.244   &8e-03&     0.251&1e-05\\
$\mu$&  $\sim0$&$N/A$&$\sim0$&$N/A$&$\sim0$&$N/A$\\ \bottomrule
 \end{tabular}\label{tbl:snsPos2}
\end{center}
\end{table}
\begin{table}[h]
\begin{center}
\caption{{Sensitivity indices $S_{i,m}$ and residuals of velocity QoIs for Test Case 2}}
\begin{tabular}{*{7}{c}}
\cmidrule(lr){2-7}
& \multicolumn{6}{c}{Quantities of Interest} \\ 
\midrule Random & \multicolumn{2}{c}{$\hat{x}$}& \multicolumn{2}{c}{$\hat{y}$}& \multicolumn{2}{c}{$\hat{z}$} \\ 
\cmidrule(lr){2-7}
Inputs&PCE&Resid.&PCE&Resid.&PCE&Resid.\\\midrule  
$x$&   4.30e-03&6e-05&5.93e-03&1e-05&5.99e-03&3e-05\\
$y$&    0.291&2e-04&     0.284&1e-03&     0.277&2e-04\\
$z$&     0.239&1e-04&     0.244&1e-03&     0.236&7e-04\\
$\dot{x}$&    0.0386&7e-03&    0.0428&6e-04&    0.0426&1e-04\\
$\dot{y}$&     0.207&5e-03&     0.194&6e-04&     0.192&4e-04\\
$\dot{z}$&     0.257&5e-03     &0.251&2e-04&     0.246&1e-03\\
$\mu$&  $\sim0$&$N/A$&$\sim0$&$N/A$&$\sim0$&$N/A$\\ \bottomrule
 \end{tabular}\label{tbl:snsVel2}
\end{center}
\end{table}
\subsection{Test Case 3}

\begin{figure}[h] 
 \centerline{\includegraphics[width = \textwidth]{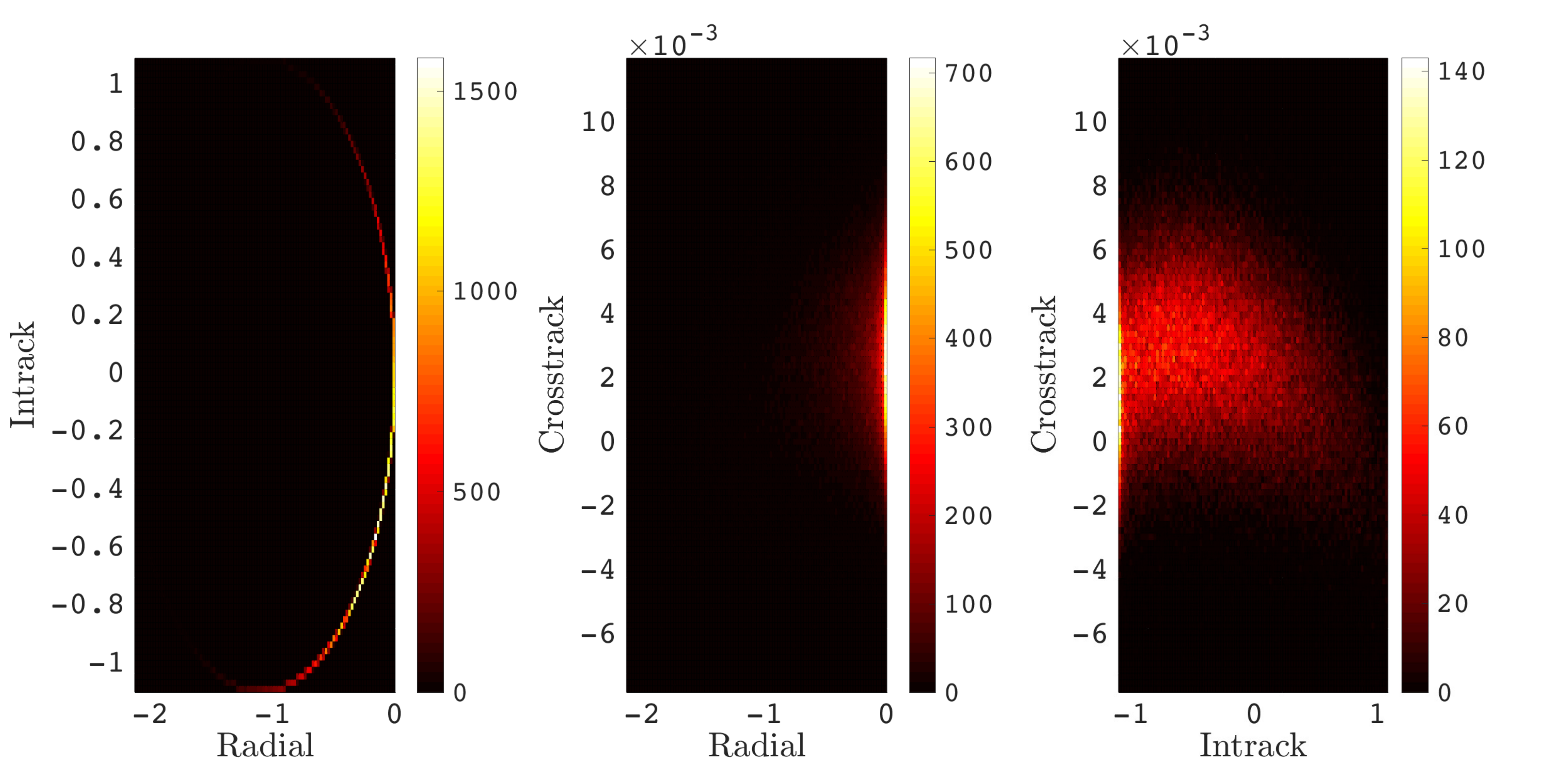}}
\caption{{MC results for Test Case 3 plotted in the RIC coordinates. Note that the colorbars illustrate object count.}}
 \label{fig:RIC_equi}
\end{figure}

For Test Case 3, a scenario presented by~\cite{horwood} is considered. For this, the initial conditions, random inputs and relevant standard deviations are found in Table~\ref{tbl:std_equi}. The distribution of 100,000 MC samples in RIC coordinates and the non-Gaussian distributions can be seen in Fig.~\ref{fig:RIC_equi}.

\begin{figure}[h] 
 \centerline{\includegraphics[width = \textwidth]{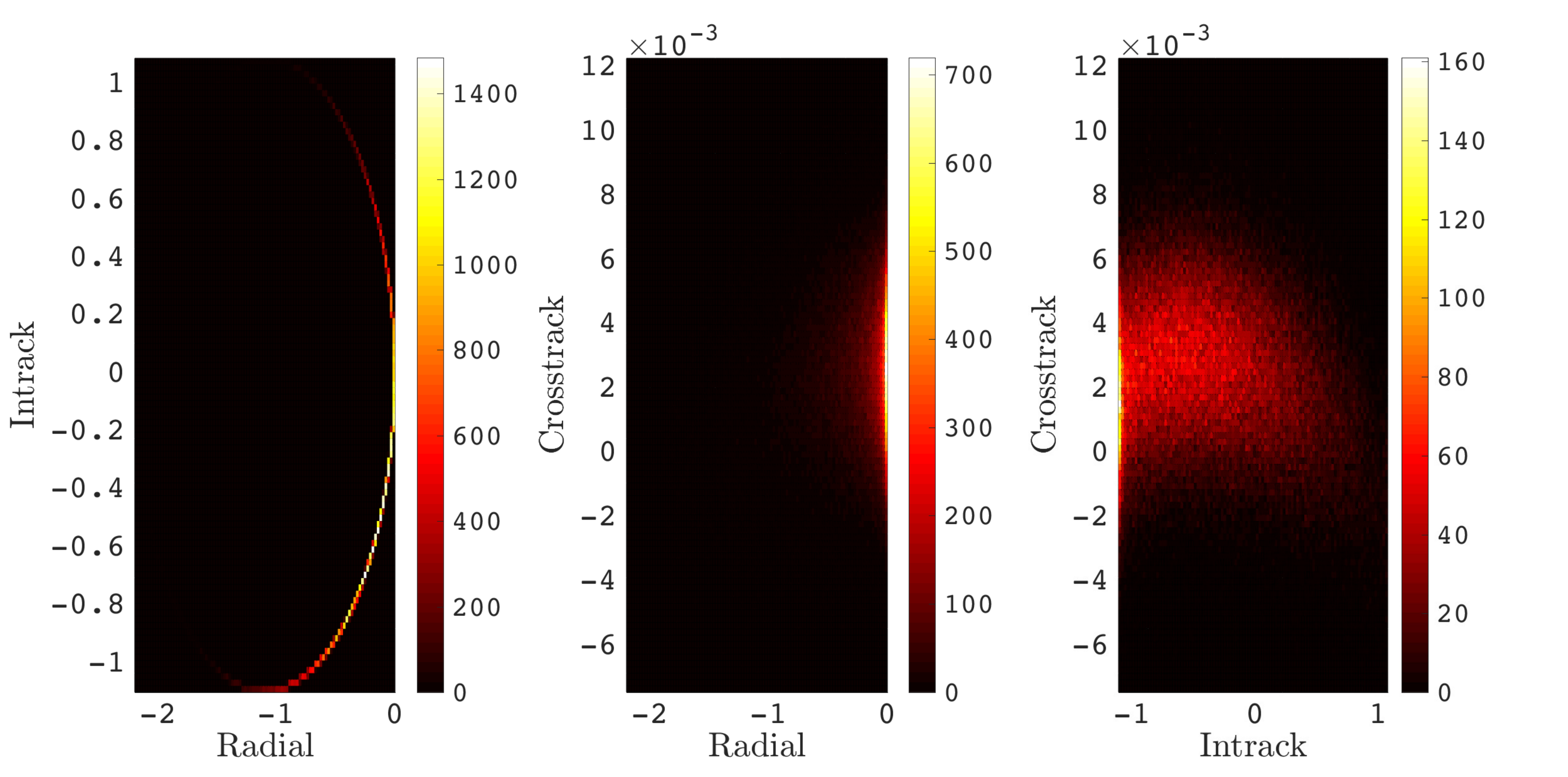}}
\caption{{SR results for Test Case 3 plotted in the RIC coordinates. Note that the colorbars illustrate object count.}}
 \label{fig:RIC_equiSR}
\end{figure}

\begin{figure}[h] 
 \centerline{\includegraphics[width = \textwidth]{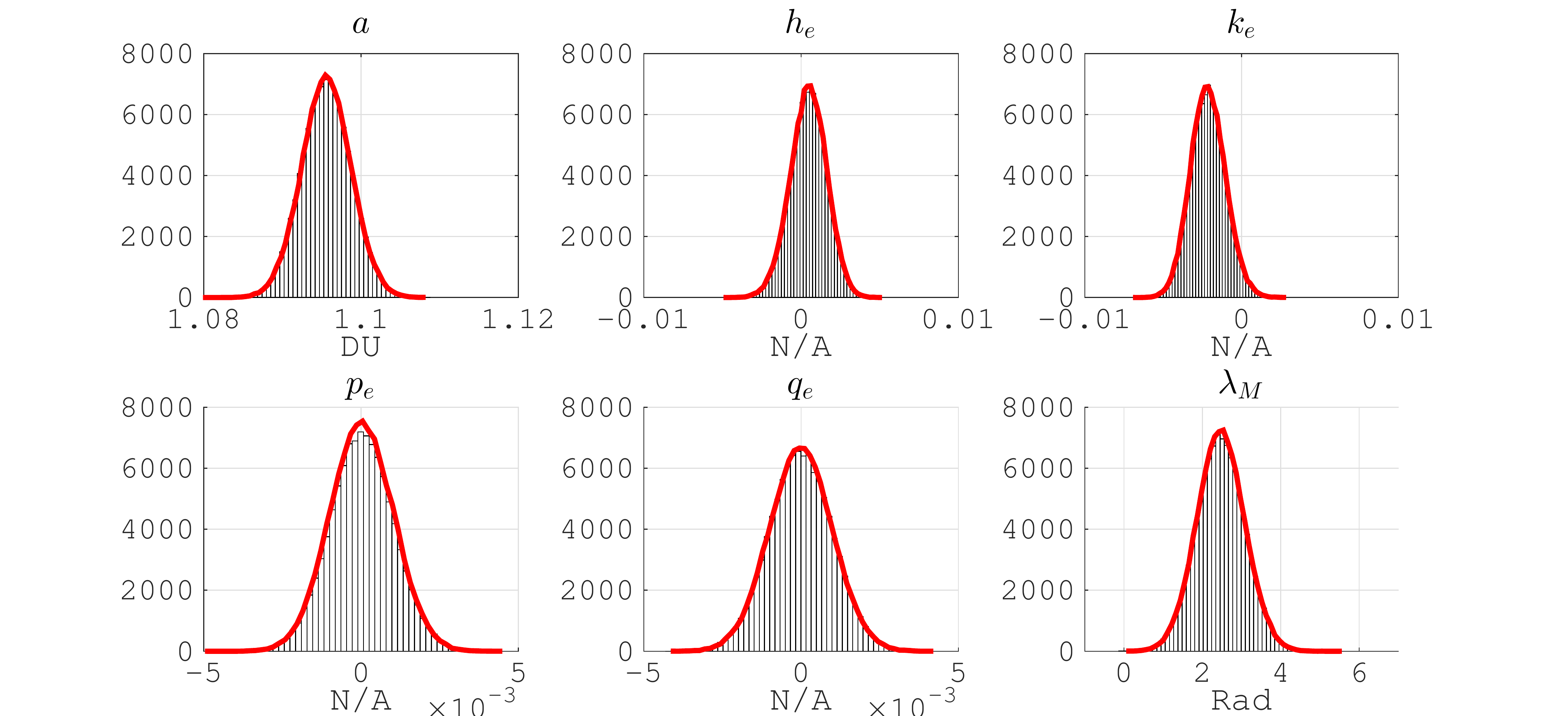}}
\caption{Histograms of SR results for quantities of interest in Test Case 3.}
 \label{fig:hist_equi}
\end{figure}

Using $N = 200$ training samples, $r = 6$, $P = 4$ and $\delta = 10^{-6}$, an SR surrogate is estimated for the six equinoctial elements. { Figure~\ref{fig:RIC_equiSR} presents the 100,000 SR realizations in the RIC frame, which compares well to Figure~\ref{fig:RIC_equi}. The results of the moment analysis, and therefore the quantitative fit, can be seen in Table~\ref{tbl:stats_3}. Three digits or more of precision is shown by the relative residuals for mean and standard deviation. 
Table~\ref{tbl:val_3} includes the validation results for Test Case 3. The table presents RMS values of the relative residuals for 40 independent random input vectors, as well as the RMS values of the MC-based validation samples for comparison.} In addition to this, Figure~\ref{fig:hist_equi} shows the SR result as a histogram plotted alongside a 100,000 MC result. Qualitatively, it can be seen that the SR solution captures the distributions well.
%
\begin{table}[h]
\begin{center}
\caption{Agreement between SR- and PC-based mean and STD for Test Case 3}
\begin{tabular}{lllllll}\toprule   & Ref. Mean&Ref. STD &Rel. Mean&Rel. STD \\ 
\midrule 
$a$                  &1.095586 (DU) &3.1368e-03 (DU)&2.6e-07&7.6e-05\\ 
$h_e$                &4.673e-04     &1.15e-03       &7.8e-04&2.2e-03\\ 
$k_e$                &-2.130e-03    &1.115e-03      &1.5e-04&7.2e-04\\ 
$p_e$                &1.85e-05      &1.0005e-03     &4.2e-03&4.6e-05\\ 
$q_e$                &-7.05e-06     &9.9948e-04     &4.7e-03&1.3e-05\\
$\lambda_\mathcal{M}$&2.46646 (rad)&0.606994 (rad) &1.1e-06&2.2e-06\\ \bottomrule
 \end{tabular}\label{tbl:stats_3}
\end{center}
\end{table}
\begin{table}[h]
\begin{center}
\caption{{Residual RMS of 40 SR- and MC-based validation samples for Test Case 3}}
\begin{tabular}{llll}\toprule  &MC Sample RMS&Residual RMS&Units\\ \midrule 
$a$  &1.09614&4.4e-05&(DU)\\ 
$h_e$&1.3e-03&3.8e-04&(N/A)\\ 
$k_e$&2.3e-03&4.1e-04&(N/A)\\ 
$p_e$&9.1e-03&2.0e-05&(N/A)\\ 
$q_e$&1.02e-03&1.8e-05&(N/A)\\
$\lambda_\mathcal{M}$&2.4063&4.8e-04&(rad)\\ \bottomrule
 \end{tabular}\label{tbl:val_3}
\end{center}
\end{table}

\begin{figure}[h] 
 \centerline{\includegraphics[width = \textwidth]{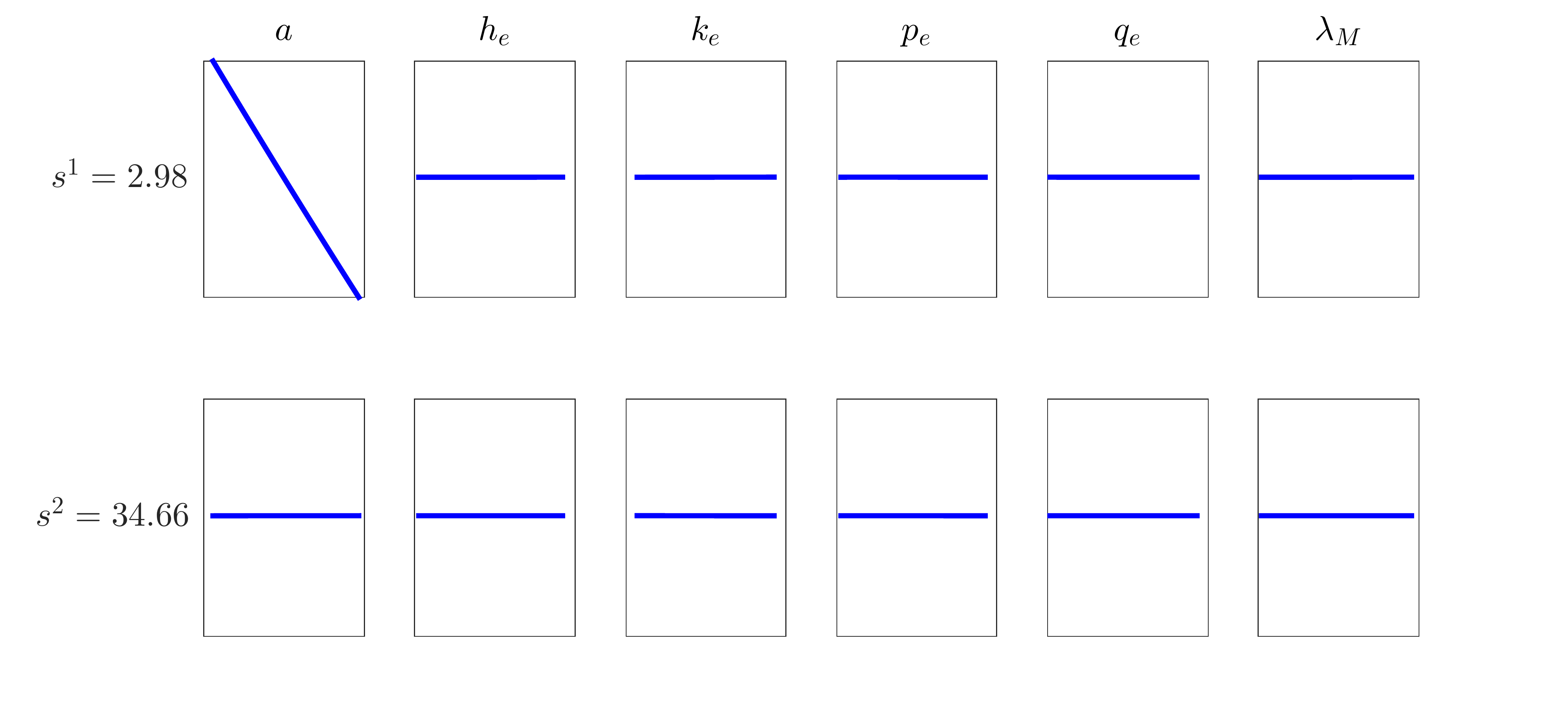}}
\caption{Plots of the absolute values of factors $u_i^l(\xi_i)$ for Test Case 3}
 \label{fig:u_eq}
\end{figure}

In~\cite{horwood}, it is stated that uncertainty in the semimajor axis $a$ has the largest effect on the uncertainty of the final PDF. At a qualitative glance, Figure~\ref{fig:u_eq} appears to agree with this conclusion. The figure is derived similarly to that of Fig.~\ref{fig:u_d20}, with the appropriate $\bm{c}^l_i$, ${\bm{u}}_0^l$, and polynomial bases being used in lieu of those in Test Case 2. Higher ranks have been omitted from the figure due to the lack of variability with respect to the scale of the image. Depicting the univariate functions for each dimension and rank, the figure shows that the first rank of dimension $a$ contains significantly more variability than any other univariate function. 

\begin{figure}[h] 
 \centerline{\includegraphics[width = \textwidth]{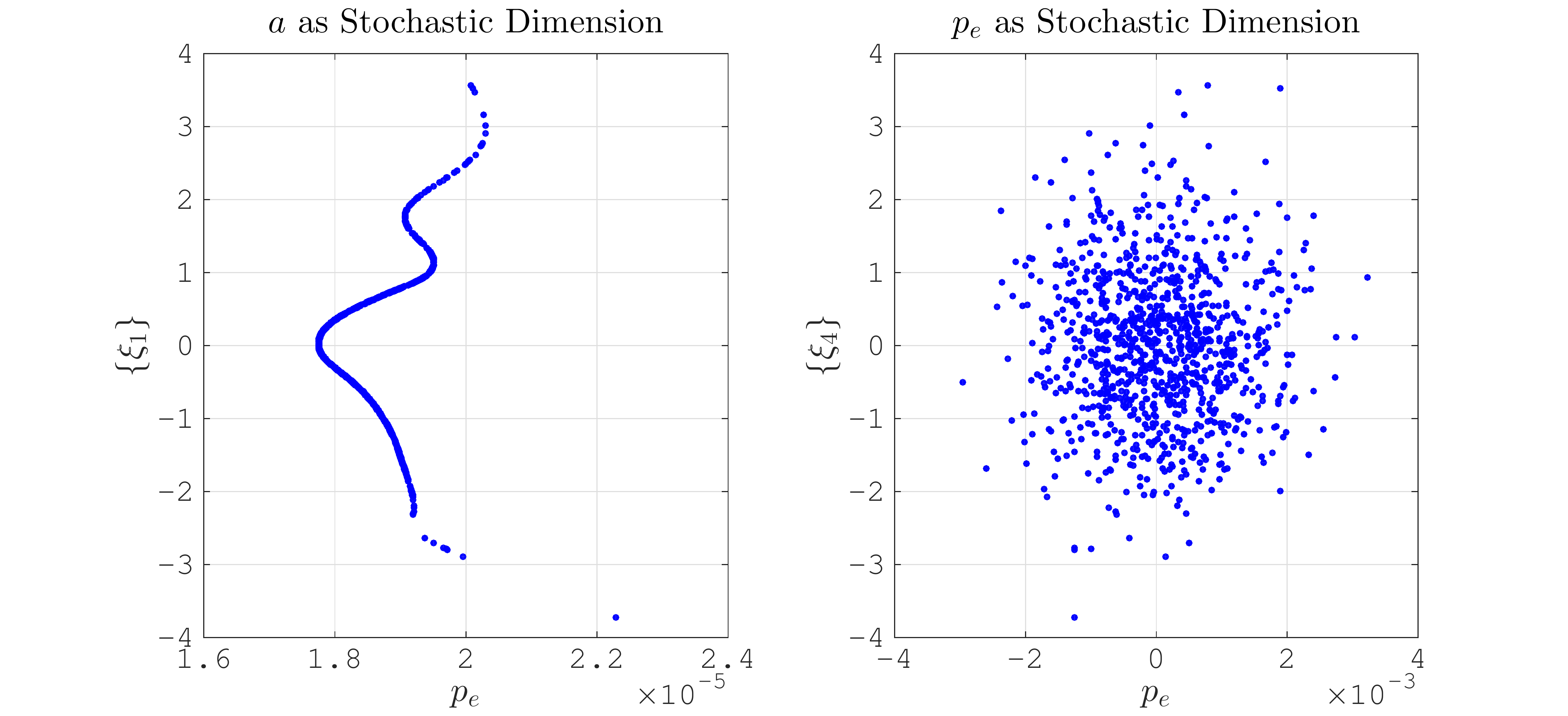}}
\caption{Plots of $p_e$ propagated with uncertainty in $a$ or $p_e$}
 \label{fig:p_e}
\end{figure}

{In order to quantitatively validate these conclusions, a sensitivity analysis is applied to the results of Test Case 3 and compared to a PC baseline. The results of this analysis can be found in Tables~\ref{tbl:sns13}~and~\ref{tbl:sns23}. As in Test Case 2, the PC baseline is converged to six decimal places. Therefore, some values have been stated to be approximately 0.} In \cite{horwood}, it is deduced that uncertainty in $a$ is the most important contribution to equinoctial element variance. In particular, the variance of $\lambda_\mathcal{M}$ relies largely on uncertainty in $a$. Tables~\ref{tbl:sns13}~and~\ref{tbl:sns23} show that variance in $a$ and $\lambda_\mathcal{M}$ does indeed rely mostly on uncertainty in $a$. However, the Sobol indices present a more detailed analysis. The results are indicative of a relatively independent system, where the variability of each QoI is most affected by the variability of its corresponding random input. 
\begin{table}[h]
\begin{center}
\caption{{Sensitivity indices $S_{i,m}$ and residuals of $a$, $h_e$ and $k_e$ for Test Case 3}}
\begin{tabular}{*{7}{c}}
\cmidrule(lr){2-7}
& \multicolumn{6}{c}{Quantities of Interest} \\ 
\midrule Random & \multicolumn{2}{c}{$a$}& \multicolumn{2}{c}{$h_e$}& \multicolumn{2}{c}{$k_e$} \\ 
\cmidrule(lr){2-7}
Inputs&PCE&Resid.&PCE&Resid.&PCE&Resid.\\\midrule  
  $a$                   &0.999    &1e-03   &0.249    &5e-03&0.200&3e-03\\
$h_e$                   &$\sim 0$&$N/A$    &0.723    &8e-03&0.0297&4e-04\\
$k_e$                   &2e-06&2e-07    &0.0267    &1e-03&0.769&3e-03\\
$p_e$                   &$\sim 0$&$N/A$&$\sim 0$&$N/A$    &$\sim 0$   &$N/A$\\
$q_e$                   &$\sim 0$&$N/A$&$\sim 0$&$N/A$    &$\sim 0$    &$N/A$\\
$\lambda_{\mathcal{M}}$ &$\sim 0$&$N/A$&$\sim 0$&$N/A$&$\sim 0$&$N/A$\\ \bottomrule
\end{tabular}\label{tbl:sns13}
\end{center}
\end{table}
\begin{table}[h]
\begin{center}
\caption{{Sensitivity indices $S_{i,m}$ and residuals of $p_e$, $q_e$ and $\lambda_{\mathcal{M}}$ for Test Case 3}}
\begin{tabular}{*{7}{c}}
\cmidrule(lr){2-7}
& \multicolumn{6}{c}{Quantities of Interest} \\ 
\midrule Random & \multicolumn{2}{c}{$p_e$}& \multicolumn{2}{c}{$q_e$}& \multicolumn{2}{c}{$\lambda_{\mathcal{M}}$} \\ 
\cmidrule(lr){2-7}
Inputs&PCE&Resid.&PCE&Resid.&PCE&Resid.\\\midrule  
  $a$                   &4e-06   &3e-06    &4e-06  &4e-06&0.999&1e-03\\
$h_e$                   &$\sim 0$&$N/A$   &$\sim 0$  &$N/A$&$\sim 0$&$N/A$\\
$k_e$                   &$\sim 0$&$N/A$    &$\sim 0$ &$N/A$&$\sim 0$&$N/A$\\
$p_e$                   &0.963 &2e-04       &0.0364  &3e-07&$\sim 0$&$N/A$\\
$q_e$                   &0.0364 &3e-05     &0.963   &1e-03&$\sim 0$&$N/A$\\
$\lambda_{\mathcal{M}}$ &$\sim 0$&$N/A$    &$\sim 0$&$N/A$&$\sim 0$&$N/A$\\ \bottomrule
\end{tabular}\label{tbl:sns23}
\end{center}
\end{table}
Figure~\ref{fig:p_e} illustrates the variability of the QoI $p_e$ propagated with uncertainty only in $a$ or uncertainty in $p_e$. The uncertainty used in this analysis is taken from the respective values in Table~\ref{tbl:std_equi}. Therefore, the dependence of the variability of $p_e$ with respect to uncertainty in $a$ and $p_e$ can be compared. As illustrated, the effect of uncertainty in $p_e$ is two orders of magnitude larger than the variability introduced by $a$. This independent behavior repeats for $h_e$, $k_e$ and $q_e$ and explains the relatively high rank, $r = 6$, needed for solution convergence. 

\FloatBarrier

\section{Conclusion}\label{sec:cnclsn}
{Separated Representations (SR) is a polynomial surrogate method of propagating uncertainty that approximates the mapping of random variables to the quantities of interest. The theoretical number of samples required for generating an SR approximation is linear with respect to stochastic dimension, which is seen in the presented results when increasing the number of random inputs from 6 to 20. Generating a surrogate approximation of the function of a space object's uncertain state does not assume posterior Gaussian distributions. Subsequently, such an approximation enables a tractable means of global sensitivity analysis when considering the uncertain state of a space object. Results shown in this paper demonstrate that the SR-based Sobol indices agree with those generated via a polynomial chaos expansion with a known accuracy. A sensitivity analysis for an equinoctial case presented in previous work expands on the hypothesis that the posterior uncertainty depends most on semimajor axis uncertainty. With the exception of semimajor axis and the angle quantity, each quantity of interest varies most as a function of its random input.}

\section{Acknowlegements}
The material for the work by Marc Balducci is provided by the NSTRF fellowship, NASA Grant NNX15AP41H.

This material is based upon work of Alireza Doostan supported by the U.S. Department of Energy Office of Science, Office of Advanced Scientific Computing Research, under Award Number DE-SC0006402, and NSF grants DMS-1228359 and CMMI-1454601.

\section{Appendix} \label{sec:appndx}
\begin{table}[h]
\begin{center}
\caption{Low degree Stokes coefficients}
\begin{tabular}{lllc}\toprule & Value&STD\\ \midrule
C$_{2,0}$&-4.8416e-04&6.1e-11\\
C$_{2,2}$&2.4393e-06&3.1e-11\\
C$_{3,0}$&9.5721e-07&1.1e-11\\
C$_{3,1}$&2.0304e-06&1.6e-11\\
C$_{3,2}$&9.0479e-07&2.2e-11\\
C$_{3,3}$&7.2127e-07&2.6e-11\\
C$_{4,0}$&5.3999e-07&8.2e-12\\
S$_{2,2}$&-1.4002e-06&3.1e-11\\
S$_{3,1}$&2.4820e-07&1.6e-11\\
S$_{3,2}$&-6.1898e-07&2.2e-11\\
S$_{3,3}$& 1.4143e-06&2.6e-11
\\\bottomrule
\end{tabular}\label{tbl:stokes}
\end{center}
\end{table}

Note that $f_r$ is a \textit{retrograde factor}, where it is +1 for all direct orbits and $-1$ for nearly retrograde orbits~\citep{vallado}.~\begin{equation}
h_e = e \sin{\omega + f_r \Omega}
\end{equation}
\begin{equation}
k_e = e \cos{\omega + f_r \Omega}
\end{equation}
\begin{equation}
p_e = \frac{\sin{i}\sin{\Omega}}{1 + \cos^{f_r}{i}}
\end{equation}
\begin{equation}
q_e = \frac{\sin{i}\cos{\Omega}}{1 + \cos^{f_r}{i}}
\end{equation}
\begin{equation}
\lambda_\mathcal{M} = \mathcal{M} + \omega + f_r \Omega
\end{equation}

\FloatBarrier
\bibliographystyle{spbasic}
\bibliography{SR_bib}

\begin{thebibliography}{49}
\providecommand{\natexlab}[1]{#1}
\providecommand{\url}[1]{{#1}}
\providecommand{\urlprefix}{URL }
\expandafter\ifx\csname urlstyle\endcsname\relax
  \providecommand{\doi}[1]{DOI~\discretionary{}{}{}#1}\else
  \providecommand{\doi}{DOI~\discretionary{}{}{}\begingroup
  \urlstyle{rm}\Url}\fi
\providecommand{\eprint}[2][]{\url{#2}}

\bibitem[{Ammar et~al(2008)Ammar, Chinesta, and Joyot}]{ammar}
Ammar A, Chinesta F, Joyot P (2008) The nanometric and micrometric scales of
  the structure and mechanics of materials revisited: An introduction to the
  challenges of fully deterministic numerical descriptions. International
  Journal for Multiscale Computational Engineering 6(3):191--213,
  \doi{10.1615/IntJMultCompEng.v6.i3.20}

\bibitem[{Askey and Arthur(1985)}]{askey85}
Askey RA, Arthur WJ (1985) Some basic hypergeometric orthogonal polynomials
  that generalize Jacobi polynomials, vol 319. AMS, Providence RI

\bibitem[{Balducci et~al(2013)Balducci, Jones, and Doostan}]{balducci}
Balducci M, Jones BA, Doostan A (2013) Orbit uncertainty propagation with
  separated representations. AAS/AIAA Astrodynamics Specialist Conference
  Hilton Head, SC, August 11-15

\bibitem[{Beylkin and Mohlenkamp(2005)}]{beylkin05}
Beylkin G, Mohlenkamp MJ (2005) Algorithms for numerical analysis in high
  dimensions. SIAM Journal on Scientific Computing 26(6):2133--2159,
  \doi{10.1137/040604959}

\bibitem[{Beylkin et~al(2009)Beylkin, Garcke, and Mohlenkamp}]{beylkin}
Beylkin G, Garcke J, Mohlenkamp MJ (2009) Multivariate regression and machine
  learning with sums of separable functions. SIAM Journal on Scientific
  Computing 31(3):1840--1857, \doi{10.1137/070710524}

\bibitem[{Blatman and Sudret(2010)}]{sudret}
Blatman G, Sudret B (2010) Efficient computation of global sensitivity indices
  using sparse polynomial chaos expansions. Reliability Engineering and System
  Safety 95:1216--1229, \doi{10.1016/j.ress.2010.06.015}

\bibitem[{Chevreuil et~al(2013)Chevreuil, Lebrun, Nouy, and Rai}]{chevreuil13}
Chevreuil M, Lebrun R, Nouy A, Rai P (2013) A least-squares method for sparse
  low rank approximation of multivariate functions. arXiv preprint
  arXiv:13050030

\bibitem[{Chinesta et~al(2011)Chinesta, Ladeveze, and
  Cueto}]{chinesta2011short}
Chinesta F, Ladeveze P, Cueto E (2011) A short review on model order reduction
  based on proper generalized decomposition. Archives of Computational Methods
  in Engineering 18(4):395--404

\bibitem[{Cohen et~al(2010)Cohen, DeVore, and Schwab}]{Cohen10a}
Cohen A, DeVore R, Schwab C (2010) Convergence rates of best n-term galerkin
  approximations for a class of elliptic spdes. Foundations of Computational
  Mathematics 10(6):615--646

\bibitem[{DeMars et~al(2013)DeMars, Bishop, and Jah}]{demars1}
DeMars KJ, Bishop RH, Jah MK (2013) Entropy-based approach for uncertainty
  propagation of nonlinear dynamical systems. Journal of Guidance, Control, and
  Dynamics 36(4):1047--1057, \doi{10.2514/1.58987}

\bibitem[{DeMars et~al(2014)DeMars, Cheng, and Jah}]{demars2}
DeMars KJ, Cheng Y, Jah MK (2014) Making best use of model evaluations to
  compute sensitivity indices. Journal of Guidance, Control, and Dynamics
  37(3):979--984, \doi{10.2514/1.62308}

\bibitem[{Doostan and Iaccarino(2009)}]{Doostan09}
Doostan A, Iaccarino G (2009) A least-squares approximation of partial
  differential equations with high-dimensional random inputs. Journal of
  Computational Physics 228(12):4332--4345

\bibitem[{Doostan et~al(2007)Doostan, Iaccarino, and Etemadi}]{Doostan07b}
Doostan A, Iaccarino G, Etemadi N (2007) A least-squares approximation of
  high-dimensional uncertain systems. Tech. Rep. Annual Research Brief, Center
  for Turbulence Research, Stanford University

\bibitem[{Doostan et~al(2013)Doostan, Validi, and Iaccarino}]{doostan}
Doostan A, Validi A, Iaccarino G (2013) Non-intrusive low-rank separated
  approximation of high-dimensional stochastic models. Computation Methods in
  Applied Mechanical Engineering 263:42--55, \doi{10.1016/j.cma.2013.04.003}

\bibitem[{Fann et~al(2004)Fann, Beylkin, Harrison, and Jordan}]{beylkin04}
Fann G, Beylkin G, Harrison R, Jordan K (2004) Singular operators in
  multiwavelet bases. IBM Journal of Research and Development 48(2):161--171,
  \doi{10.1147/rd.482.0161}

\bibitem[{Forrester et~al(2008)Forrester, Sobester, and
  Keane}]{forrester2008engineering}
Forrester A, Sobester A, Keane A (2008) Engineering design via surrogate
  modelling: a practical guide. John Wiley \& Sons

\bibitem[{Friedman et~al(2001)Friedman, Hastie, and
  Tibshirani}]{friedman2001elements}
Friedman J, Hastie T, Tibshirani R (2001) The elements of statistical learning,
  vol~1. Springer series in statistics Springer, Berlin

\bibitem[{Fujimoto et~al(2012)Fujimoto, Scheeres, and Alfriend}]{fujimoto}
Fujimoto K, Scheeres DJ, Alfriend KT (2012) Analytical nonlinear propagation of
  uncertainty in the two-body problem. Journal of Guidance, Control, and
  Dynamics 35(2):497--509, \doi{10.2514/1.54385}

\bibitem[{Ghanem and Spanos(1991)}]{ghanem}
Ghanem R, Spanos P (1991) Stochastic Finite Elements: A Spectral Approach.
  Springer-Verlag, New York

\bibitem[{Hadigol et~al(2014)Hadigol, Doostan, Matthies, and
  Niekamp}]{hadigol14}
Hadigol M, Doostan A, Matthies HG, Niekamp R (2014) Partitioned treatment of
  uncertainty in coupled domain problems: A separated representation approach.
  Computer Methods in Applied Mechanics and Engineering 274:103--124,
  \doi{10.1016/j.cma.2014.02.004}

\bibitem[{Hampton and Doostan(2015)}]{hampton15b}
Hampton J, Doostan A (2015) Coherence motivated sampling and convergence
  analysis of least squares polynomial chaos regression. Computer Methods in
  Applied Mechanics and Engineering pp 73--97,
  \doi{doi:10.1016/j.cma.2015.02.006}

\bibitem[{Hansen and Schwab(2012)}]{Hansen12}
Hansen M, Schwab C (2012) Analytic regularity and nonlinear approximation of a
  class of parametric semilinear elliptic pdes. Mathematische Nachrichten

\bibitem[{Harrison et~al(2004)Harrison, Fann, Yanai, Gan, and
  Beylkin}]{harrison}
Harrison RJ, Fann GI, Yanai T, Gan Z, Beylkin G (2004) Multiresolution quantum
  chemistry: Basic theory and initial applications. Journal of Chemical Physics
  121(23), \doi{10.1063/1.1791051}

\bibitem[{Hoang and Schwab(2013)}]{Hoang13}
Hoang VH, Schwab C (2013) Sparse tensor galerkin discretization of parametric
  and random parabolic pdes---analytic regularity and generalized polynomial
  chaos approximation. SIAM Journal on Mathematical Analysis 45(5):3050--3083

\bibitem[{Horwood et~al(2011)Horwood, Aragon, and Poore}]{horwood}
Horwood JT, Aragon ND, Poore AB (2011) Gaussian sum filters for space
  surveillance: Theory and simulations. Journal of Guidance, Control, and
  Dynamics 34(6):1839--1851, \doi{10.2514/1.53793}

\bibitem[{Jones and Doostan(2013)}]{jones_doostan}
Jones BA, Doostan A (2013) Satellite collision probability estimation using
  polynomial chaos expansions. Advances in Space Research 52(11):1860--1875,
  \doi{10.1016/j.asr.2013.08.027}

\bibitem[{Jones et~al(2014{\natexlab{a}})Jones, Doostan, and Born}]{jones_mlt}
Jones BA, Doostan A, Born G (2014{\natexlab{a}}) Nonlinear propagation of orbit
  uncertainty using non-intrusive polynomial chaos. Journal of Guidance,
  Control, and Dynamics \doi{10.2514/1.57599}, in press, available from$\colon$
  URL
  http$\colon\/\/$ccar.colorado.edu$\/$bajones$\/$files$\/$jones\_2014a.pdf.

\bibitem[{Jones et~al(2014{\natexlab{b}})Jones, Parrish, and Doostan}]{jones}
Jones BA, Parrish N, Doostan A (2014{\natexlab{b}}) Post-maneuver collision
  probability estimation using sparse polynomial chaos expansions. Journal of
  Guidance, Control, and Dynamics 36(2):430--444, \doi{10.2514/1.G000595}

\bibitem[{Junkins et~al(1996)Junkins, Akella, and Alfriend}]{junkins}
Junkins JL, Akella MR, Alfriend KT (1996) Non-gaussian error propagation in
  orbital mechanics. Journal of Astronautical Sciences 44(4):541--563

\bibitem[{Khoromskij and Schwab(2010)}]{khoromskij10}
Khoromskij BN, Schwab C (2010) Tensor-structured galerkin approximation of
  parametric and stochastic elliptic pdes. SIAM Journal on Scientific Computing
  33:364--385, \doi{10.1137/100785715}

\bibitem[{Kolda and Bader(2009)}]{Kolda09b}
Kolda TG, Bader BW (2009) Tensor decompositions and applications. SIAM Review
  51(3):455--500

\bibitem[{Majji et~al(2008)Majji, Junkins, and Turner}]{majji}
Majji M, Junkins J, Turner J (2008) A high order method for estimation of
  dynamic systems. The Journal of the Astronautical Sciences 56(3):401--440,
  \doi{10.1007/BF03256560}

\bibitem[{Nielsen et~al(2012)Nielsen, Alfriend, Bloomfield, Emmert, Miller,
  Guo, and et~al.}]{nrc_2012}
Nielsen P, Alfriend K, Bloomfield M, Emmert J, Miller J, Guo Y, et~al (2012)
  Continuing Kepler's Quest: Assessing Air Force Space Command's Astrodynamic
  Standards. The National Academies Press, Washington DC

\bibitem[{Nouy(2010)}]{nouy10}
Nouy A (2010) Proper generalized decompositions and separated representations
  for the numerical solution of high dimensional stochastic problems. Archives
  of Computational Methods in Engineering 17:403--434,
  \doi{10.1007/s11831-010-9054-1}

\bibitem[{Park and Scheeres(2006)}]{park}
Park RS, Scheeres DJ (2006) Nonlinear mapping of gaussian statistics: Theory
  and applications to spacecraft trajectory design. Journal of Guidance,
  Control, and Dynamics 29(6):1367--1375, \doi{10.2514/1.20177}

\bibitem[{Quadrelli et~al(2015)Quadrelli, Wood, Riedel, McHenry, Aung,
  Cangahuala, Volpe, Beauchamp, and Cutts}]{quadrelli}
Quadrelli MB, Wood LJ, Riedel JE, McHenry MC, Aung M, Cangahuala LA, Volpe RA,
  Beauchamp PM, Cutts JA (2015) Guidance, navigation, and control technology
  assessment for future planetary science missions. Journal of Guidance,
  Control, and Dynamics 38(7):1165--1186, \doi{10.2514/1.G000525}

\bibitem[{Reynolds et~al(2015)Reynolds, Doostan, and Beylkin}]{reynolds15}
Reynolds M, Doostan A, Beylkin G (2015) Randomized alternating least squares
  for canonical tensor decompositions: Application to a pde with random data.
  arXiv preprint arXiv:151001398

\bibitem[{Russell(2012)}]{ryan}
Russell R (2012) Survey of spacecraft trajectory design in strongly perturbed
  environments. Journal of Guidance, Control, and Dynamics 35(3):705--720,
  \doi{10.2514/1.56813}

\bibitem[{Sabol et~al(2011)Sabol, Binz, Segerman, Roe, and
  Schumacher~Jr}]{sabol11}
Sabol C, Binz C, Segerman A, Roe K, Schumacher~Jr PW (2011) Probability of
  collision with special perturbations dynamics using the monte carlo method.
  In: AAS/AIAA Astrodynamics Specialist Conference, Girdwood, AK

\bibitem[{Saltelli(2002)}]{saltellis}
Saltelli A (2002) Making best use of model evaluations to compute sensitivity
  indices. Computer Physics Communications 145():280–297,
  \doi{10.1016/S0010-4655(02)00280-1}

\bibitem[{Saltelli et~al(2004)Saltelli, Tarantola, Campolongo, and
  Ratto}]{saltelli}
Saltelli A, Tarantola S, Campolongo F, Ratto M (2004) Sensitivity Analysis in
  Practice: A guide to assessing scientific models. Wiley, Hoboken, NJ

\bibitem[{Schutz et~al(2004)Schutz, Tapley, and Born}]{born}
Schutz B, Tapley B, Born GH (2004) Statistical orbit determination. Academic
  Press

\bibitem[{Smith(2013)}]{rsmithUQ}
Smith R (2013) Uncertainty Quantification: Theory, Implementation, and
  Applications. SIAM-Society for Industrial and Applied Mathematics

\bibitem[{Sobol(2001)}]{sobol}
Sobol I (2001) Global sensitivity indices for nonlinear mathematical models and
  their monte carlo estimates. Mathematics and Computers in Simulation
  55:271--280

\bibitem[{Sun and Kumar(2015)}]{sun}
Sun Y, Kumar M (2015) Uncertainty propagation in orbital mechanics via tensor
  decomposition. Celestial Mechanics and Dynamical Astronomy pp 1--26,
  \doi{10.1007/s10569-015-9662-z}

\bibitem[{Tamellini et~al(2014)Tamellini, Le~Maitre, and Nouy}]{tamellini14}
Tamellini L, Le~Maitre O, Nouy A (2014) Model reduction based on proper
  generalized decomposition for the stochastic steady incompressible
  navier-stokes equations. SIAM Journal on Scientific Computing
  36(3):A1089--A1117, \doi{10.1137/120878999}

\bibitem[{Tapley et~al(2005)Tapley, Ries, Bettadpur, Chambers, Cheng, Condi,
  Gunter, Kang, P.Nagel, Pastor, Pekker, S.Poole, and Wang}]{tapley}
Tapley B, Ries J, Bettadpur S, Chambers D, Cheng M, Condi F, Gunter B, Kang Z,
  PNagel, Pastor R, Pekker T, SPoole, Wang F (2005) Ggm02 - an improved earth
  gravity field model from grace. Journal of Geodesy
  \doi{10.1007/s00190-005-0480-z}

\bibitem[{Vallado(2007)}]{vallado}
Vallado D (2007) Fundamentals of Astrodynamics and Applications, 3rd edn,
  Microcosm Press, Hawthorne, CA, chap 8.6, p 562

\bibitem[{Xiu(2010)}]{xiu10a}
Xiu D (2010) Numerical Methods for Stochastic Computations: A Spectral Method
  Approach. Princeton University Press

\end{thebibliography}

\end{document}